\documentclass{article}

\usepackage{PRIMEarxiv}

\usepackage[utf8]{inputenc} 
\usepackage[T1]{fontenc}    
\usepackage{hyperref}       
\usepackage{url}            
\usepackage{booktabs}       
\usepackage{amsfonts}       
\usepackage{nicefrac}       
\usepackage{microtype}      
\usepackage{lipsum}
\usepackage{fancyhdr}       
\usepackage{graphicx}       
\graphicspath{{media/}}     
\usepackage{caption}  
\usepackage{subcaption} 
\usepackage[utf8]{inputenc}
\usepackage{amsfonts}
\usepackage{amsmath}
\usepackage{booktabs}
\usepackage{multirow}
\usepackage{makecell}
\usepackage{array}
\title{Self-Supervised Learning for WiFi CSI-Based Human Activity Recognition: A Systematic Study
\thanks{This work was done when Ke Xu was at I2R, A*STAR.} 
}

\author{
  Ke Xu \\
  Coventry University \\
  Coventry, UK\\
  \texttt{xuk16@uni.coventry.ac.uk} \\
   \And
  Jiangtao Wang \\
  Coventry University \\
  Coventry, UK\\
  \texttt{jiangtao.wang@coventry.ac.uk} \\
  \AND
  Hongyuan Zhu \\
  Institute for Infocomm Research (I$^2$R), A*STAR \\
  Singapore \\
  \texttt{zhuh@i2r.a-star.edu.sg} \\
  \And
  Dingchang Zheng \\
  Coventry University \\
  Coventry, UK\\
  \texttt{dingchang.zheng@coventry.ac.uk} \\
}

\begin{document}
\maketitle

\begin{abstract}
Recently, with the advancement of the Internet of Things (IoT), WiFi CSI-based HAR has gained increasing attention from academic and industry communities.
By integrating the deep learning technology with CSI-based HAR, researchers achieve state-of-the-art performance without the need of expert knowledge.
However, the scarcity of labeled CSI data remains the most prominent challenge when applying deep learning models in the context of CSI-based HAR due to the privacy and incomprehensibility of CSI-based HAR data.
On the other hand, SSL has emerged as a promising approach for learning meaningful representations from data without heavy reliance on labeled examples.
Therefore, considerable efforts have been made to address the challenge of insufficient data in deep learning by leveraging SSL algorithms.
In this paper, we undertake a comprehensive inventory and analysis of the potential held by different categories of SSL algorithms, including those that have been previously studied and those that have not yet been explored, within the field.
We provide an in-depth investigation of SSL algorithms in the context of WiFi CSI-based HAR.
We evaluate four categories of SSL algorithms using three publicly available CSI HAR datasets, each encompassing different tasks and environmental settings.
To ensure relevance to real-world applications, we design performance metrics that align with specific requirements.
Furthermore, our experimental findings uncover several limitations and blind spots in existing work, highlighting the barriers that need to be addressed before SSL can be effectively deployed in real-world WiFi-based HAR applications. Our results also serve as a practical guideline for industry practitioners and provide valuable insights for future research endeavors in this field.
\end{abstract}

\keywords{WiFi \and Channel State Information \and Self-Supervised Learning \and human activity recognition}

\section{Introduction}
\label{sec:1.intro}
With the proliferation of the Internet of Things (IoT), the accessibility of WiFi devices has become increasingly ubiquitous.
Furthermore, thanks to the research on WiFi channel state information (CSI) perception technology for Human Activity Recognition (HAR) tasks, such as activity recognition \cite{wang2014eyes,yousefi2017survey,jiang2018towards,xue2020deepmv}, human identification \cite{zhang2021device,deng2022gaitfi,chen2022rfcam},fall detection \cite{zhang2015anti,wang2016wifall,palipana2018falldefi}, gesture recognition \cite{pu2013whole,zhang2021widar3,niu2021understanding} and sign language recognition \cite{ma2018signfi}, WiFi has been transformed from mere WiFi routers to multi-functional devices with sensing capabilities.
Leveraging the widespread availability of WiFi routers, the use of WiFi in indoor environments is device-free, resulting in a cost-effective solution.
Moreover, compared to wearable sensors and cameras, WiFi-based HAR is non-intrusive and robust against changes in illumination.
Accordingly, WiFi-based HAR is acknowledged by academic communities as an important technology in establishing the Artificial Intelligence of Things (AIoT) \cite{zhang2022practical}.
Recently, the emergence of deep learning has enabled models to achieve state-of-the-art performance without the need for complex preprocessing operations that incorporate expert knowledge \cite{xiao2019csigan,li2021two,xu2022dual},such as Doppler Frequency Shift and Time-of-Flight \cite{qian2018widar2}.

Despite the advantages of deep learning approaches for WiFi sensing based HAR tasks, they still face many challenges in real-world applications.
One of the most eminent challenges is the limited availability of labeled data.
Specifically, existing powerful WiFi based HAR deep learning models are primarily based on the supervised paradigm and require a large amount of annotated data.
However, acquiring a sufficient quantity of labeled data for model training is notorious cumbersome and expensive, because the raw CSI data is incomprehensible to humans.
This characteristic precludes practitioners from collecting labeled CSI data via post-labeling, which is often used in computer vison community.
Post-labeling, which differs from the approach of collecting and annotated data simultaneously, collects data followed by labeling. This significantly restricts the size of the collected labeled CSI data.
Ideally, researchers expect CSI-based HAR algorithm can train the model with limited amount of data, while demonstrating excellent capability of generalization across different environment setting.

Indeed, utilizing unlabeled data to facilitate the model training process is a straightforward way to tackle the issue of model training in the presence of scarce data.
From the perspective of machine learning, learning from unlabeled data can be formulated as unsupervised learning \cite{bengio2012unsupervised}.
Recently, as a branch of unsupervised learning, Self-Supervised Learning (SSL) has achieved great success in various computer vision and natural language processing tasks.
It has drawn massive attention for its good performance in terms of data efficiency and generalization ability.
At a high-level, the idea behind the SSL is as follows: (1)Semi-automatically extract labels from data itself. (2)Predict part of the data from any other part.
Compare to other unsupervised learning algorithms, SSL has the same paradigm as supervised learning, making it easily applicable to many existing supervised learning scenarios with minor modifications \cite{liu2021self}.

In the field of CSI-based HAR, SSL is still in a nascent stage.
Although some work has been done in this area \cite{yang2022autofi,liu2021contrastive,9141293,zhao2023self,lau2021self,saeed2021sense}, there is currently a lack of a systematic study on the feasibility of SSL for CSI-based HAR and a benchmark for fair comparison of different algorithms.
We urge researchers should carefully analyze the advantages, limitations and assumption when applying SSL to the CSI HAR tasks. 
In addition, we call for the establishment of a standardized benchmark, as in the computer vision community, to promote algorithmic research.

This paper represents an initial exploration and in-depth investigation into the feasibility of employing SSL methods in the domain of CSI-based HAR. 
To comprehensively evaluate the potential of SSL in this context, our assessment framework is established around three main pillars, as follows:
\begin{itemize}
    \item Comparison of embedding quality in SSL: We evaluate and compare the quality of embeddings produced by different SSL algorithms. This examination aims to quantify the effectiveness of SSL in learning meaningful representations from CSI data, which is crucial for subsequent HAR tasks.
    \item Robustness of SSL algorithms to domain shift of data: We assess the robustness of SSL algorithms when faced with domain shifts, where the pre-training and testing data come from different distributions or environments. Understanding the performance of SSL algorithms in scenarios with domain shifts is vital for their practical deployment in real-world applications, as it ensures their adaptability to varying conditions.
    \item Robustness of SSL algorithms to data scarcity: We analyze the data efficiency of SSL algorithms, specifically their ability to achieve strong performance with limited data. By evaluating the SSL algorithms' performance under varying amounts of data, we gain insights into their potential for reducing the reliance on large datasets, which is particularly valuable in situations where labeled data is scarce or expensive to acquire.
\end{itemize}

To the best of our knowledge, no prior work has extensively investigated and analyzed the development of SSL algorithms specifically for CSI-based HAR tasks.
As such, this work makes a unique and original contribution to the CSI-based HAR literature and aims to provide clear guidelines for practitioners or researchers to choose appropriate SSL algorithms in practice when facing different scenarios. 
The main contributions of our work are:
\begin{itemize}
    \item We conduct a comprehensive and systematic study to evaluate the performance of state-of-the-art SSL algorithms on three CSI HAR datasets collected from diverse settings. 
    \item Building upon the theoretical foundations of SSL in the machine learning literature, we assess four categories of SSL solutions within the context of CSI-based HAR.
    \item We propose performance metrics for evaluating SSL algorithms in CSI HAR, considering three key perspectives that align with real-world application scenarios.
    \item Though utilizing the largest publicly available dataset in the field of CSI HAR, Widar \cite{zhang2021widar3}, for SSL experiments under various settings, we gained numerous novel insights and were able to provide practical guidelines for researchers and practitioners in the CSI HAR field. 
\end{itemize}

\begin{figure}
     \centering
         \includegraphics[width=\linewidth]{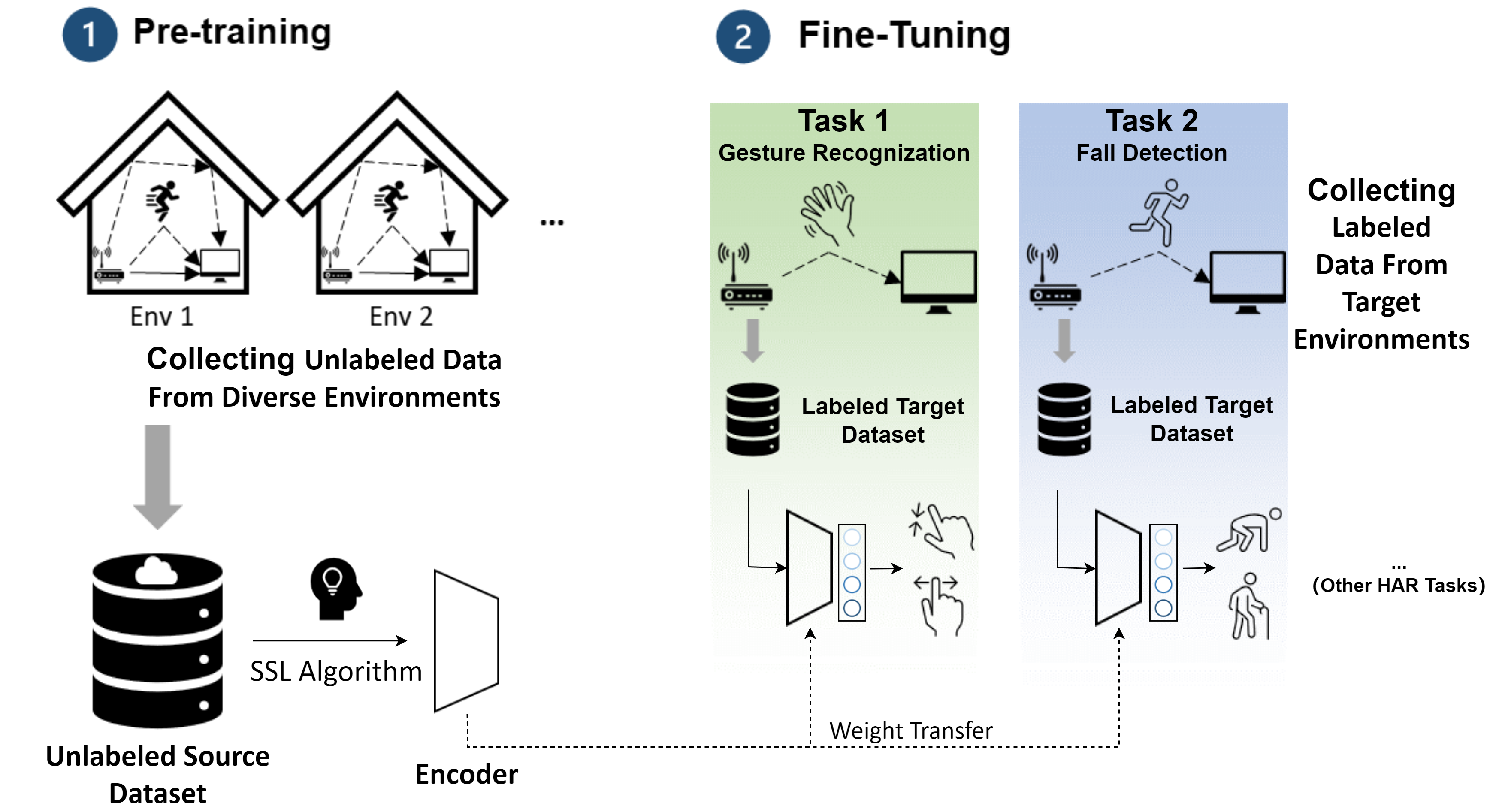}
        \caption{The self-supervised learning paradigm consists of two stages: (1) unlabeled self-supervised pre-training and (2) labeled supervised fine-tuning.}
        \label{fig:SSLflowchart}
\end{figure}

\section{Related Work}

\subsection{Deep Learning on CSI based HAR} 
Researchers have explored both handcrafted features and deep learning methods for WiFi-based HAR.
Deep learning-based methods are standing at the forefront due to their automatic feature extraction with full annotation.
These methods can be categorized based on the type of deep model utilized. 
multi-layer perceptron is one of the classic architectures used in this area, with studies such as \cite{liu2017wicount} and \cite{zhang2018crosssense} utilizing multi-layer perceptron for people counting and gesture recognition, respectively. 
Long short-term memory (LSTM) has also been used by researchers such as \cite{yousefi2017survey}, \cite{chen2018wifi}, and \cite{ma2021location} to extract temporal information from CSI data. 
Convolutional Neural Network (CNN), a commonly used model in computer vision, has been utilized in several studies.
For example, \cite{jiang2018towards} proposed an adversarial network for activity recognition using CNN. 
In addition, \cite{xiao2020deepseg} proposed a CNN-based algorithm to transform activity segmentation into classification problems, and \cite{yang2022robustsense} proposed a learning framework using CNN to achieve consistent predictions regardless of whether an attack is present in the input. 
Additionally, THAT \cite{li2021two} proposed a Transformer-based model based on the characteristics of CSI data and achieved state-of-the-art performance.
Some recent studies have combined handcrafted features with deep learning algorithms to gain excellent performance. 
Basing on domain knowledge, Widar \cite{zhang2021widar3} propose a theoretically domain-independent feature body-coordinate velocity profile (BVP). 
On this basis, they achieve one-fit-all performance using a hybrid CNN-RNN model that can generalize to different data domains with only one-time training.

\subsection{Self-Supervised Learning on CSI based HAR}
While the aforementioned deep learning approaches have exhibited high recognition accuracy in HAR, they continue to encounter issues with inadequate labeled data due to their reliance on the supervised learning paradigm.
To address this issue, SSL has emerged as a promising area that can leverage massive amounts of unlabeled data to learn transferable and semantically meaningful embeddings through auxiliary tasks, without requiring manual annotations. 
In the domain of WiFi-based HAR, SSL has garnered increasing attention, with several contrastive-based SSL methods achieving performances that are comparable to supervised baselines \cite{chen2020simple}. 
In \cite{saeed2020federated}, a contrastive learning paradigm is used for extract semantic feature from two views of raw signal and Scalgram.
\cite{liu2021semi} combines semi-supervised learning and contrastive learning to further reduces the labeling burden of HAR applications.
In addition to existing time-domain data augmentation function, researchers also pay attention to develop frequency-domain data augmentation during self-supervised training process \cite{liu2021contrastive}.
Furthermore, \cite{saeed2021sense} introduced a generalized framework named Sense and Learn, which can learn high-level features using several original auxiliary tasks, without requiring tedious labeling processes.

We have listed all these work and the category of SSL method they used in Table \ref{table:related work}. 
Although these works have achieved remarkable success, they are typically limited to the simple framework of SimCLR \cite{chen2020simple}, while ignoring a wider and more comprehensive spectrum of SSL algorithms. Hence, in this study, we aim to systematically explore the performance of SSL frameworks on CSI HAR tasks.
\begin{table}[!htbp]
\begin{tabular}{c|cccc}
\toprule
Method    &Cluster Discrimination  &Instance Discrimination &Relation Prediction &Autoencoder\\
\midrule 
SiFall\cite{ji2022sifall}              &  &  &  &\checkmark     \\
\cite{zhao2023self}                    &  &  &\checkmark  &     \\
BTS\cite{shen2022semi}                 &  &\checkmark  &  &     \\
\cite{koupai2022self}                  &  &  &  &\checkmark     \\
RF-URL\cite{song2022rf}                &  &\checkmark  &  &     \\
AutoFi\cite{yang2022autofi}            &  &\checkmark  &  &     \\
SemiC-HAR\cite{liu2021semi}            &  &\checkmark  &  &     \\
STF-CSL\cite{liu2021contrastive}       &  &\checkmark  &  &     \\
\cite{saeed2021sense}                  &  &  &\checkmark  &     \\
\cite{lau2021self}                     &  &\checkmark  &  &     \\
\cite{saeed2020federated}              &  &\checkmark  &  &     \\
Ours        &\checkmark &\checkmark &\checkmark &\checkmark     \\

\bottomrule
\end{tabular}
\caption{The table includes the categories of SSL algorithms utilized by previous methods, as well as the categories of SSL algorithms investigated in our study.}
\label{table:related work}
\end{table}


\section{Preliminaries and Research}
\subsection{Motivation}
\label{sec:3.1.moti}
As elucidated in Section \ref{sec:1.intro}, CSI-based HAR holds substantial promise for diverse applications, including motion detection for elderly individuals residing alone. 
This bears particular relevance in the context of the global aging trend, as major economies worldwide are undergoing a transition towards aging societies \cite{han2020aging}. 
Consequently, enhancing the quality of life for solitary elderly individuals has emerged as a pressing concern. Moreover, the proliferation of smart home devices has instigated interest in achieving AIOT within residential environments, opening up a promising avenue for exploration.

In this context, we can envision an application scenario involving an elderly man named Jack, who lives alone and has limited mobility. By treating smart devices as WiFi access points, an activity recognition system can be developed to enhance Jack's daily routines within his home, providing convenience and assistance. For instance, the system could enable Jack to interact with smart furniture through gesture recognition, control indoor temperature, or even operate a robotic sweeper using gestures. Importantly, the system can also detect and respond to accidental falls by alerting and calling for help.

SSL, employing the "pre-train then fine-tune" paradigm, demonstrates promise in training an effective HAR model utilizing cost-effective unlabeled CSI datasets. However, before applying SSL in this scenario, several pertinent questions necessitate attention.

\textbf{Q1: Whether SSL can extract high-quality features so that it can generalize across diverse tasks?}
Specifically, the feature learned from the pretext task in the pre-training stage should perform well with a simple linear classifier across diverse HAR tasks, including but not limited to gesture recognition.

\textbf{Q2: How does SSL perform in the presence of data distribution discrepancies between pre-training and fine-tuning datasets?}
For example, if Jack is not represented in the collected unlabeled dataset, can SSL still yield satisfactory performance? Additionally, the issue of disparate environments between the pre-training dataset and the target environment warrants consideration. Can the SSL-trained model proficiently operate in an environment it has never encountered during the pre-training stage? Furthermore, it is also essential to investigate the impact of changes in receiver positions on the model's performance.

\textbf{Q3: Can SSL algorithms effectively handle low-data scenarios?}
The acquisition of labeled target datasets can be prohibitively expensive, underscoring the need to reduce reliance on labeled data. However, determining the minimum amount of labeled data required for the CSI HAR model to achieve satisfactory performance remains an open question. Furthermore, striking a balance between the cost of acquiring the unlabeled dataset and the efficacy of pre-training is crucial for practitioners. Therefore, it becomes imperative to explore the impact of the unlabeled dataset's size on the performance of SSL-based models.
In Section \ref{sec:4.Exp}, we structure our analysis and experiments based on these three dimensions.

\subsection{CSI Data}
In summary, the CSI describes the propagation of wireless signals in the physical environment between the WiFi transmitter and receiver. Employing Multiple-Input Multiple-Output and Orthogonal Frequency Division Multiplexing techniques in the physical layer of the IEEE 802.11n/ac standards enhances data capacity and orthogonality in transmission channels impacted by multi-path propagation.

The transmitted and received signals between the transmitter and receiver can be mathematically represented by the equation $\textbf{y}_i=\textbf{H}_i \textbf{x}_i+\textbf{n}_i$, where $i \in {1, ..., S}$. Here, $\textbf{x}_i$ and $\textbf{y}_i$ represent the transmit and received signal vectors, respectively, for the $i$th subcarrier, and $\textbf{n}_i$ represents the noise vector. The channel matrix $\textbf{H}_i$ corresponds to the CSI for the $i$th subcarrier. However, human activities can modify the multipath characteristics of the wireless channel, leading to variations in the estimated channel matrix $\textbf{H}$ caused by the reflection, scattering, and diffraction of the WiFi signal due to the presence of the human body.

Researchers in the field of CSI-based HAR commonly employ three readily available CSI tools: Intel 5300 NIC \cite{halperin2011tool}, Atheros CSI Tool \cite{xie2015precise}, and Nexmon CSI Tool \cite{nexmon:project,gringoli2019free}. These tools facilitate the recording of wireless signals with varying numbers of subcarriers and bandwidths. For example, the Intel 5300 NIC captures 30 subcarriers for each antenna pair, with a bandwidth of 20MHz. The Atheros CSI Tool allows for up to 56 subcarriers with a 20MHz bandwidth and up to 114 subcarriers with a 40MHz bandwidth. Furthermore, the Nexmon CSI Tool permits the recording of 256 subcarriers using an 80MHz bandwidth. 

In our study, we utilize a dataset captured by the Intel 5300 NIC, which is the most widely used CSI tool and was the first to be released.
\subsection{SSL Techniques}
The primary focus of this study is SSL from an algorithmic perspective. The SSL paradigm consists of two stages: pre-training and fine-tuning, as depicted in Figure \ref{fig:SSLflowchart}. During the initial stage, the model is pre-trained using a large-scale unlabeled dataset. In this pre-training phase, the model learns to extract meaningful representations from the data without relying on explicit labels. In the subsequent stage, the learned encoder weights are frozen, and the model proceeds to the supervised learning stage using a labeled target dataset. In this stage, the model is fine-tuned on the labeled data, leveraging the previously learned representations for HAR. Our specific focus within the SSL framework is on four types of SSL methods: cluster discrimination, instance discrimination, relation prediction, and autoencoder. These methods exhibit promise in learning robust representations from unlabeled data and will be thoroughly investigated in our study.

\textit{Cluster Discrimination.}
Activity recognition necessitates embeddings that capture similarities among related activities. In supervised learning, this is achieved by assigning the embeddings of similar activities to the same class label, thereby reducing their distance in the embedding space. Conversely, cluster discrimination algorithms employ clustering techniques to create pseudo labels. This process assigns activities of the same category to identical clusters, guaranteeing that embeddings of similar activities are grouped together. This clustering step facilitates the recognition of activity categories by encouraging the proximity of embeddings within the cluster. A notable algorithm in the cluster discrimination category is SwAV, which incorporates an online clustering strategy to enhance computational efficiency. Moreover, SwAV utilizes multi-crop augmentation techniques to boost the algorithm's performance, enabling it to acquire more resilient representations of activities \cite{caron2020unsupervised}.

\textit{Instance Discrimination.} 
Instance discrimination is an approach in SSL that aims to minimize the distance between similar instances in the embedding space. Positive samples, which are augmented views of the same instance, are leveraged to achieve this. However, a challenge in this process is to avoid the collapse of all outputs to a constant vector, as this would result in meaningless representations. To mitigate the collapse issue, various methods have proposed different strategies. Two notable algorithms, SimCLR and MoCo, address this problem by integrating negative sampling into their approaches \cite{chen2020simple,he2020momentum}. SimCLR resolves the collapse problem by employing a large batch size during training, thereby augmenting the number of negative samples. MoCo introduces a queue mechanism and a momentum encoder to prevent collapse. It maintains a queue of past instances' features using a momentum encoder to ensure feature consistency and increases the capacity for negative samples. In contrast, BYOL and SimSiam employ a distinct approach to tackle the collapse issue \cite{grill2020bootstrap,chen2021exploring}. They employ a stop gradient and a predictor module. Specifically, one branch of the siamese network functions as the online branch, permitting gradient backpropagation. The other branch serves as the offline branch, and stop gradient is applied to restrict the flow of gradients through it. Furthermore, a predictor is utilized on the online branch to predict the output of the offline branch. This prediction task acts as a regularization mechanism, facilitating the development of meaningful representations while avoiding collapse into a constant vector.

\textit{Relation Prediction.}
Relation prediction leverages the relationships between different parts of the same input or between augmented inputs and their original versions as pretext tasks for learning meaningful embeddings. Predicting these relationships allows the model to acquire embeddings that can be applied to various downstream tasks. An example of relation prediction is the method introduced by Doersch et al. \cite{doersch2015unsupervised}, which focuses on predicting positional relations between parts. By understanding the spatial relationships among various parts within an example, the model can capture vital structural information. Another effective approach in relation prediction involves rotating different parts of an input and inferring the degree of rotation for each part \cite{gidaris2018unsupervised}. By training the model to predict the rotation angle, it can capture invariant features that are insensitive to rotation. By utilizing relation prediction techniques, models can encode rich semantic information and capture meaningful relationships between different parts of an input. This leads to enhanced representation learning and improved performance on downstream tasks.

\textit{Autoencoder.}
The auto-encoding model is built upon the fundamental principle that a good representation should be capable of reconstructing inputs from corrupted or incomplete inputs. In the field of computer vision, two specific variants of the autoencoder, namely Variational Autoencoders (VAE) and Masked Autoencoders (MAE), have demonstrated significant effectiveness. 
The VAE algorithm incorporates variational inference with an autoencoder architecture to reconstruct the input based on the calculated latent variables.
On the other hand, the MAE algorithm partitions the input into multiple patches and selectively masks a designated proportion of them. The objective is to predict the masked patches by utilizing contextual information obtained from the unmasked patches. This process can be metaphorically compared to solving a cloze problem, where missing elements are deduced from the surrounding context.

\begin{figure*}
     \centering
     \begin{subfigure}[b]{0.22\textwidth}
         \centering
         \includegraphics[width=\textwidth]{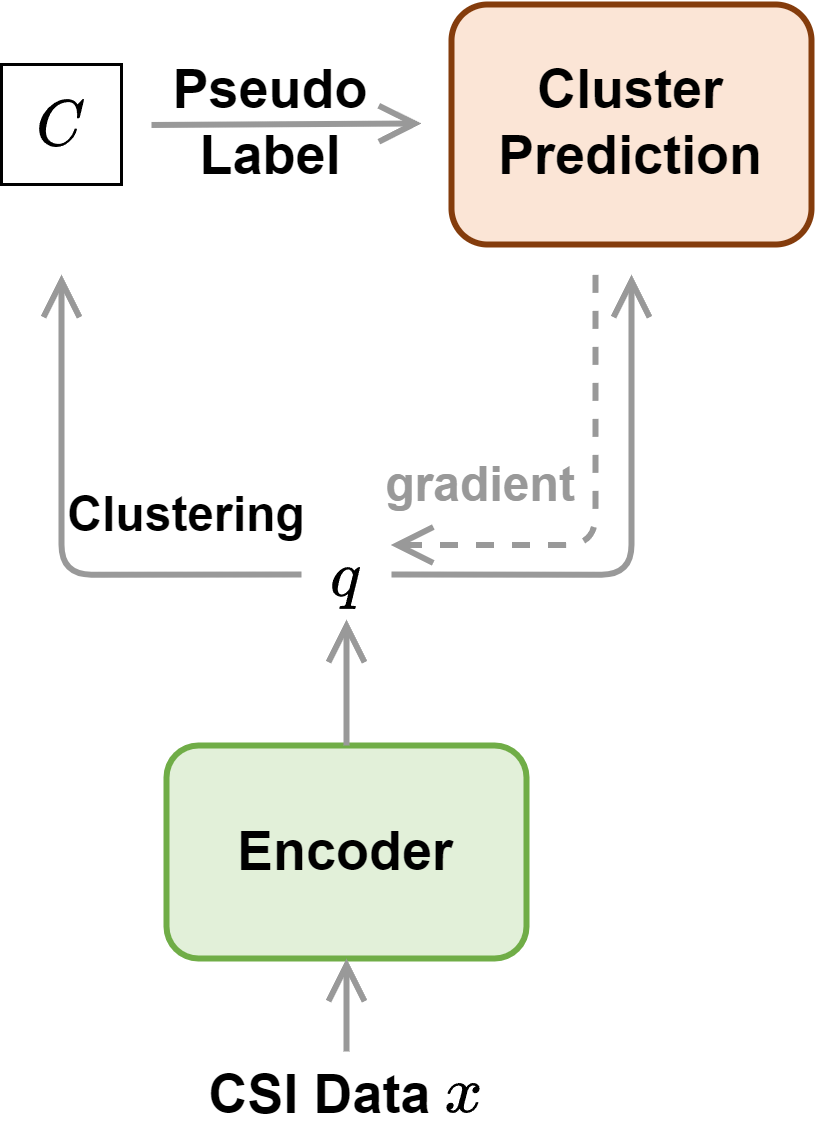}
         \caption{Cluster Discrimination}
         \label{fig:SSLARCH_a}
     \end{subfigure}
     \hfill
     \begin{subfigure}[b]{0.22\textwidth}
         \centering
         \includegraphics[width=\textwidth]{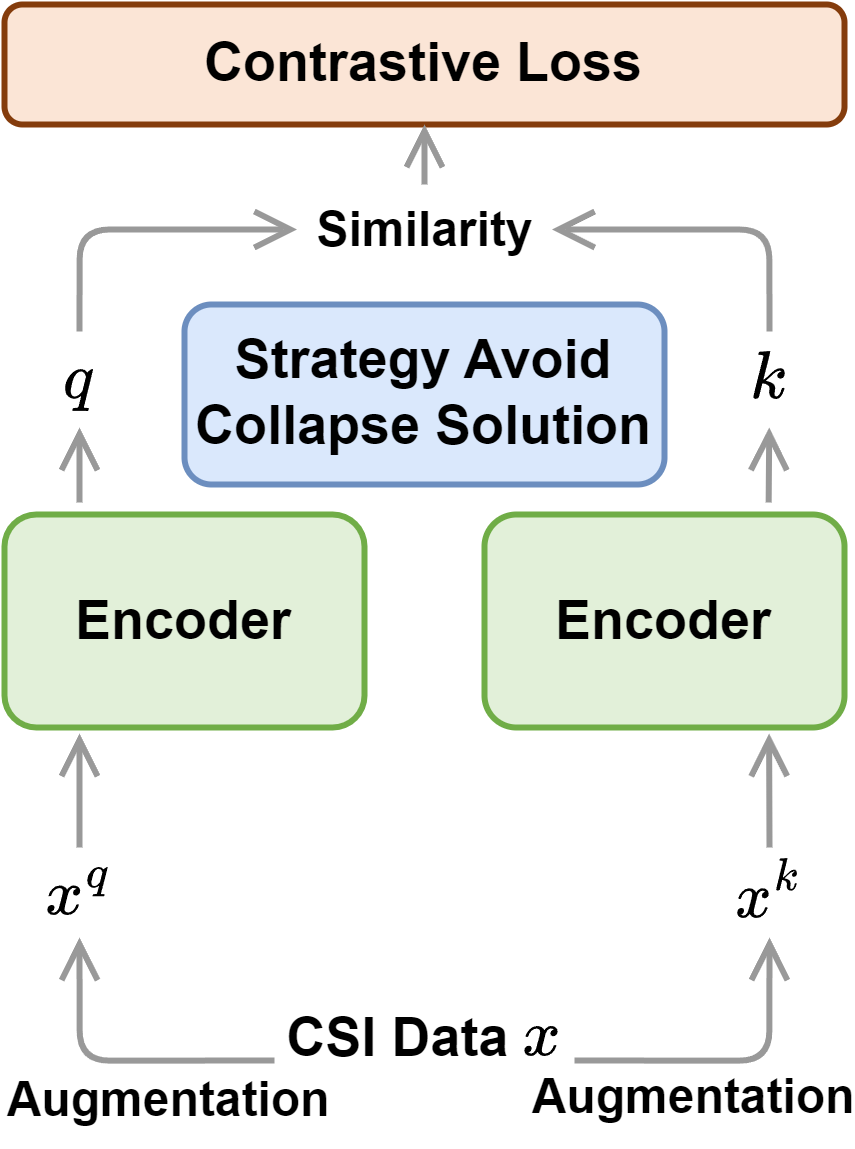}
         \caption{Instance Discrimination}
         \label{fig:SSLARCH_b}
     \end{subfigure}
     \hfill
     \begin{subfigure}[b]{0.22\textwidth}
         \centering
         \includegraphics[width=\textwidth]{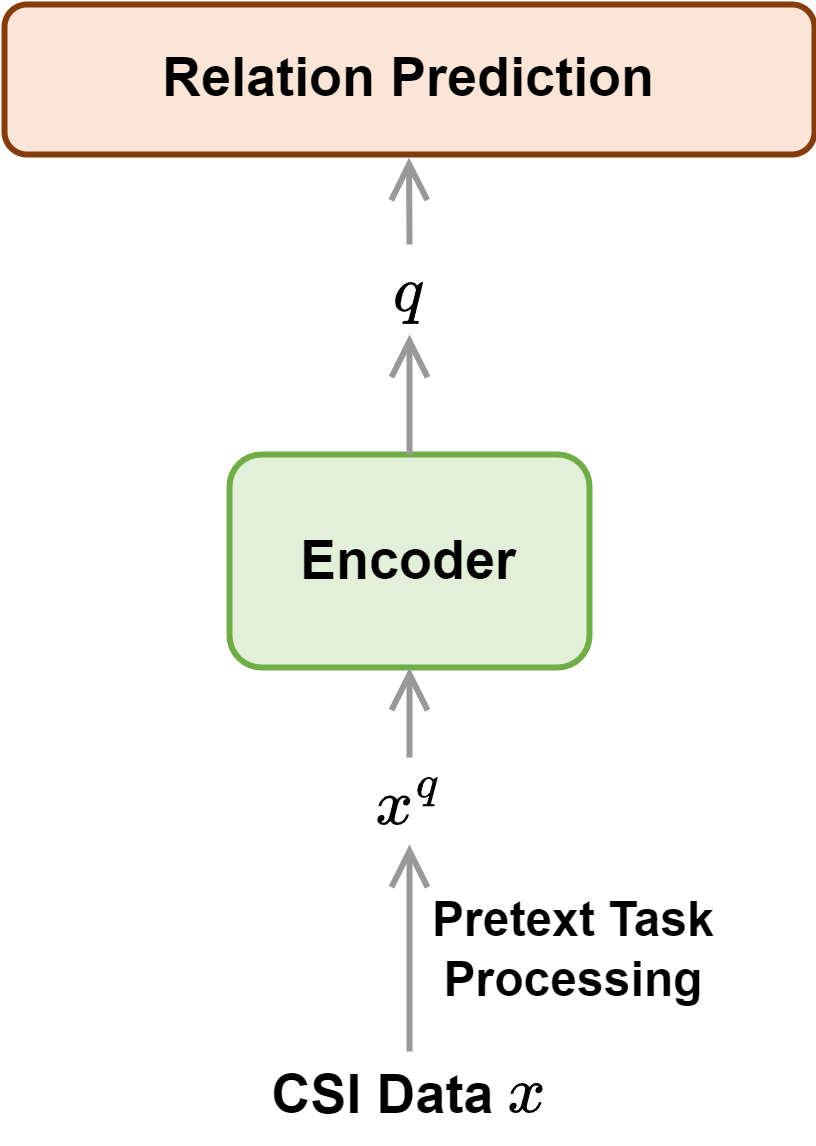}
         \caption{Relation Prediction}
         \label{fig:SSLARCH_c}
     \end{subfigure}
     \hfill
     \begin{subfigure}[b]{0.22\textwidth}
         \centering
         \includegraphics[width=\textwidth]{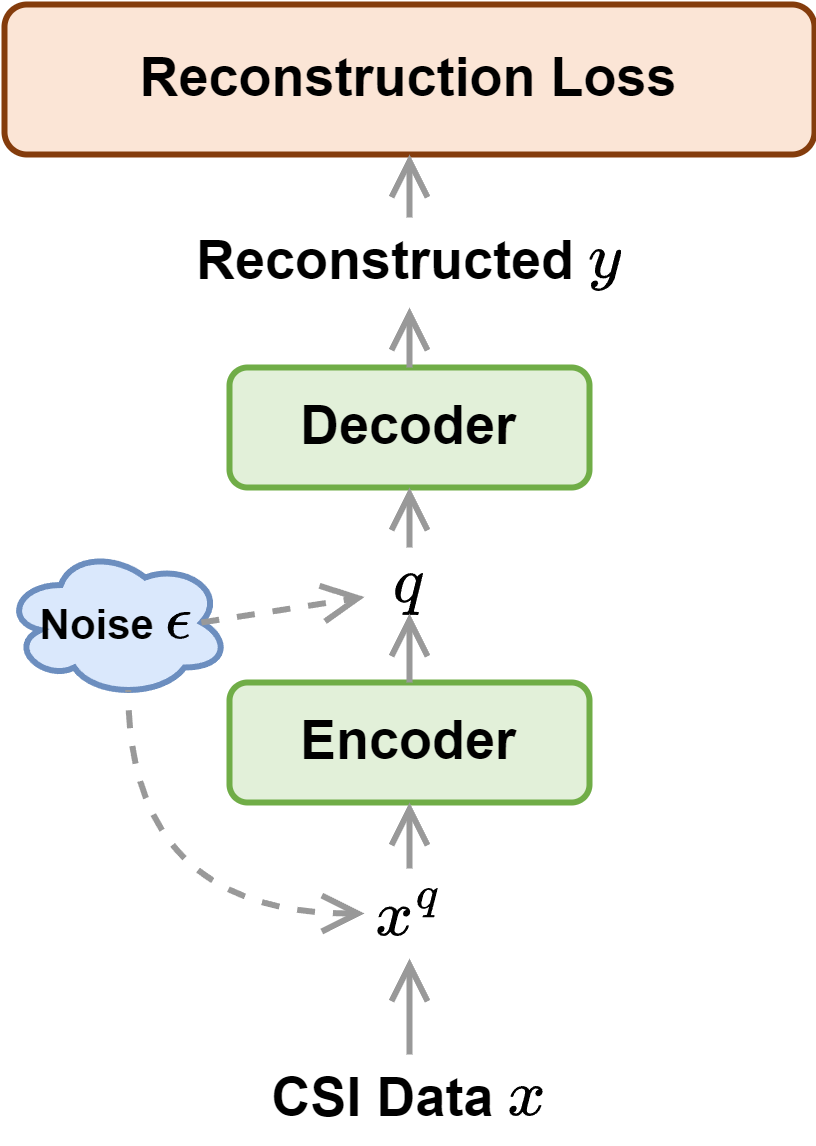}
         \caption{Autoencoder}
         \label{fig:SSLARCH_d}
     \end{subfigure}
        \caption{Different architecture pipeline of self-supervised Learning. (a) Cluster discrimination utilizes cluster to generate pseudo label for bringing similar samples together in the embedding space. (b) Instance discrimination aims to pull similar records closer in the embedding space by leveraging positive samples. (c) Relation prediction learns meaningful embeddings by predicting relationship between different parts of the same input or between augmented inputs and their original versions. (d) The autoencoder learns high-quality embeddings by reconstructing the input. To facilitate the training process, noise is added either to the input vector or to the latent variables obtained by the encoder.}
\end{figure*}

\section{Experiment}
\label{sec:4.Exp}

This section presents a systematic investigation of the feasibility of using SSL in the context of CSI-based HAR, conducted through a series of comprehensive experiments. 
As we mentioned in Section \ref{sec:3.1.moti}, our experiments is designed based on three perspectives: evaluating the quality of SSL features, assessing the robustness to distribution shifts, and analyzing the robustness to data scarcity. 
Below, we present a summary of the results and analysis.

\subsection{Setup}
To begin, we provide a comprehensive description of the dataset and outline the preprocessing steps applied to the data. Subsequently, we discuss the selection of the encoder architecture and the algorithm employed for the HAR task. Lastly, we delve into the details of the training process, which encompasses parameter selection.
\subsubsection{Dataset.}
\label{sec:5.1.data}
We select following three publicly available datasets for our experiments.

\noindent \textit{UT-HAR.} UT-HAR \cite{yousefi2017survey} is the first publicly available dataset specifically designed for HAR using CSI. 
It consists of 557 recordings collected from six participants engaging in six distinct coarse-grained activities: lying down, falling, picking up an object, running, sitting down, standing up, and walking.
The participants performed the target activities between a pair of transceivers. 
The receiver was equipped with an Intel 5300 NIC and 3 pairs of antennas, which recorded 30 subcarriers per WiFi packet. 
The data was recorded in the same room using only amplitude measurements. 
Despite its relatively small size, this dataset has been widely adopted and used in the field. Thus, we have also utilized this dataset in our experiments.

\noindent \textit{Signfi.} The Signfi dataset \cite{ma2018signfi} comprises recordings of 276 sign gestures performed by 5 participants in two distinct environments: lab and home. 
The activities were captured using a receiver equipped with an Intel 5300 NIC and 3 antennas, and each recording includes data from 30 subcarriers. 
To account for environmental variability, we used a subset of the dataset that specifically includes 2760 records recorded in the home environment. 
Each gesture in the dataset is represented by 10 records. We selected this dataset because it encompasses nearly all of the 300 most commonly used basic sign gestures in daily life.

\noindent \textit{Widar.} The Widar dataset \cite{zhang2021widar3} consists of recordings from 17 users performing 22 gesture activities in three different rooms. The dataset utilizes one transmitter and six receivers, each equipped with an Intel 5300 NIC and 3 antennas, placed at different locations. Each activity was simultaneously captured on all six receivers using 30 subcarriers, enabling the collection of multiple perspectives of the same activity. Considering each receiver's data as separate, the Widar dataset gathers a total of 271,038 data points. We selected this dataset because of its remarkable scale as the largest publicly available dataset for CSI-based HAR to date, offering a diverse range of data. However, to maintain control over variables, our subsequent experiments primarily focus on subsets of this dataset. The table below provides the detailed settings and corresponding names for each subset, as they vary in terms of the room, receiver position, and user during the data collection process.

\begin{table}[!htbp]
\begin{tabular}{c|ccccc}
\toprule
Subset Name  &Room &Receiver &User   &Act.     &Rec.\\
\midrule 
Widar\_R2     & All   & 2     & All   &22            &45170         \\
Widar\_U123R2 & All   & 2     & 1,2,3 &22            &27550         \\
Widar\_R1R2U1 & 1     & 2     & 1     &22            &10300         \\
Widar\_R1R2U2 & 1     & 2     & 2     &22            &6250         \\
Widar\_R1R2U3 & 1     & 2     & 3     &11             &4500         \\
Widar\_R2R2U1 & 2     & 2     & 1     &9              &1125         \\
Widar\_R2R2U2 & 2     & 2     & 2     &12             &2750         \\
Widar\_R2R2U3 & 2     & 2     & 3     &11             &1875         \\
Widar\_R3R2U3 & 3     & 2     & 3     &6              &750         \\
Widar\_R1R1   & 1     & 1     & All   &22            &33424         \\
Widar\_R1R2   & 1     & 2     & All   &22            &33425         \\
Widar\_R1R5   & 1     & 5     & All   &22            &33424         \\
\bottomrule
\end{tabular}
\caption{The description of subsets of Widar. The table describes the setting (including room, receiver position and user) of different subsets. The columns Act. and Rec. stand for the number of activity categories and the number of records, respectively.}
\label{table:Widarsubset}
\end{table}

\subsubsection{Data characteristics and pre-processing.}
Unless otherwise stated, the WiFi CSI records used in our experiments comprise both amplitude and phase components, resulting in a record shape of $(T,30,3,2)$, where $T$ represents the time duration of the record, 30 represents the number of subcarriers, 3 represents the number of antennas, and 2 represents the amplitude and phase components of the CSI data. 
In contrast, data that only contains either amplitude or phase has a shape of $(T,30,3)$. 
In Appendix \ref{App: different combination}, we present detailed information regarding the performance differences observed when using different combinations of CSI amplitude and phase.
To address the variability in the length of the datasets employed in our study, we implemented a preprocessing technique proposed in \cite{li2021two}. This technique involves utilizing a mean-pooling operator on adjacent time points to standardize the length of each record to a predefined time duration. We set the time length $T$ to 2000 for UT-HAR, 200 for Signfi, and 500 for Widar, ensuring that the records retain their primary information while adhering to the predefined time duration. We refrain from applying any denoising procedure prior to training, with the exception of utilizing the mean-pooling operator, which indeed filters out high-frequency noise to some degree.

\subsubsection{Algorithms}
As mentioned earlier, the primary objective of this paper is to perform a systematic investigation into the effectiveness of SSL for WiFi-based HAR tasks. 
The results of this study inherently rely on the specific SSL algorithm employed. 
Therefore, we present a detailed introduction to the SSL algorithm employed in our experiments:

\textit{SwAV} The SwAV, as proposed by Caron et al. \cite{caron2020unsupervised}, is a cluster discrimination-based SSL method. The main concept underlying SwAV is to assign different views of the same image to the same prototypes. Additionally, SwAV utilizes online clustering to expedite computations and incorporates a multi-view augmentation strategy to enhance performance.

\textit{SimCLR} The SimCLR, as proposed by Chen et al. \cite{chen2020simple}, utilizes a representation learning approach involving contrasting variously augmented views of a given input record. Negative pairs in this approach are composed of data from the same batch. Through the use of an instance discrimination-based SSL approach, SimCLR demonstrates simplicity and flexibility while generating highly effective representations for CSI data.

\textit{MoCo} The MoCo, as proposed by He et al. \cite{he2020momentum} and additionally improved by Chen et al. \cite{chen2021empirical}, represents another prominent SSL method based on instance discrimination. Similarly to SimCLR, it contrasts differently augmented views of a given input example to achieve representation learning. However, unlike SimCLR, it incorporates a memory bank, enabling it to leverage more negative pairs without requiring additional GPU memory. In Appendix \ref{App:Data Augmentation}, we present the impact of using different augmentation functions for pre-training of SwAV, SimCLR and MoCo.

\textit{Rel-Pos} The Rel-Pos, as proposed by Doersch et al. \cite{doersch2015unsupervised}, focuses on learning meaningful representations by predicting the relative position of two patches from the same image. The model creates a meaningful representation space for subsequent tasks by leveraging the spatial information obtained from the pretext task of predicting relative positions. To apply Rel-Pos to our CSI data, we consider the time duration and subcarriers as the width and height of the image, respectively.

\textit{MAE} The Masked Autoencoder (MAE), as proposed by He et al. \cite{he2022masked}, a type of Autoencoder, accomplishes this by utilizing a masked reconstruction loss function, encouraging the model to learn a sparse and informative representation of the input data. Specifically, MAE employs a binary mask that randomly zeroes out a portion of the input data and trains the model to reconstruct the original data based on the masked input. The resulting representation is both sparse and informative, rendering it suitable for diverse downstream tasks.
\begin{table}
\begin{tabular}{l|l}
\toprule
\textbf{Algorithm}         & \textbf{Type of self-supervised learning}      \\ \midrule
SwAV              & Cluster Discrimination \\ 
SimCLR            & Instance Discrimination         \\ 
MoCo              & Instance Discrimination         \\ 
Relative Position & Relation Prediction             \\ 
MAE               & Autoencoder      \\
\bottomrule
\end{tabular}
\caption{Categories of self-supervised learning algorithms used in this paper.}
\label{table:1}
\end{table}

\subsubsection{Encoder Architecture}
Besides the choice of algorithm, the selection of the architecture of the encoder significantly impacts the performance of an SSL approach. Considering our focus on SSL algorithms, we aim to use a unified encoder with different algorithms to enable a comprehensive performance comparison in this study. As stated in \cite{yang2022deep}, convolutional-based networks demonstrate better performance and computational efficiency compared to Recurrent neural networks on CSI data. In contrast, MAE solely relies on the transformer-based model, ViT \cite{dosovitskiy2020image}. Therefore, our paper specifically focuses on the following three encoder architectures:

\textit{Causal Net} Causal Net, proposed by Franceschi et al. \cite{franceschi2019unsupervised}, is a deep learning architecture developed to model causal relationships in sequential data. In contrast to traditional CNNs designed for stationary image data, causal CNNs incorporate temporal causality by restricting each output to depend solely on past inputs, disregarding future inputs. This guarantees that the model's predictions solely rely on available information at the time of prediction, making it particularly well-suited for time series data such as CSI.

\textit{ResNet} ResNet, proposed by He et al. \cite{he2016deep}, addresses the issue of degradation in deep neural networks, where the network's accuracy decreases with increasing layers. This issue was mitigated by introducing residual connections between adjacent layers in the network. By employing this approach, ResNet enables the learning of residual mappings, representing the difference between input and output of a layer, instead of direct mappings. This enables the construction of highly deep neural networks.

\textit{ViT} ViT, proposed by Dosovitskiy et al. \cite{dosovitskiy2020image}, is a network based on the Transformer architecture that employs the self-attention mechanism for image classification. In contrast to conventional convolutional networks, ViT represents the input image as a sequence of patches and flattens and embeds them using an embedding layer. Furthermore, the self-attention mechanism in ViT enables capturing global dependencies among patches, making it particularly effective at modeling long-range dependencies in data. Therefore, ViT has the potential to capture the temporal dynamics of the CSI data.

\noindent Apart from supervised learning, MoCo is compatible with all three encoder architectures. However, certain algorithms are restricted to specific encoder architecture. More specifically, SimCLR and SwAV do not support ViT. Rel-Pos is compatible only with ResNet, while MAE is compatible exclusively with ViT.

\subsubsection{Training Process}
Our experimental code was implemented using the MMSelfSup framework \cite{mmselfsup2021}, a PyTorch-based framework \cite{paszke2019pytorch} specifically designed for SSL in computer vision. 
We made necessary modifications to ensure compatibility with CSI data. 
In this section, we present the hyperparameter search spaces for all SSL approaches and provide detailed information for each method.

Since our experiments were conducted on a dataset from a distinct domain than the original papers, we extensively performed parameter search experiments to identify the optimal hyperparameter combinations. 
To establish the supervised learning baseline, we conducted a grid search to optimize the hyperparameters for training the network from scratch on the CSI dataset. 
For SSL algorithms, we optimized the hyperparameters using grid search during the pre-training stage and used the same hyperparameters as the supervised learning baseline during the fine-tuning stage. 

In our experiments, we employed the AdamW optimization algorithm and performed a hyperparameter search on the following parameters: learning rate (1e-2, 1e-3, 1e-4, 1e-5), warm-up iterations (20, 40, 60), warm-up ratios (1e-3, 1e-4, 1e-5), and batch sizes (32, 64, 128, 256). 
The values in parentheses indicate the available options for each respective hyperparameter, resulting in a total of 144 possible combinations. 

The dataset was split into training, validation, and test sets using a ratio of 0.6, 0.2, 0.2, respectively. The training set was used for model training, the validation set for selecting the best epoch, and the test set for evaluating the model performance. 

After identifying the optimal hyperparameter combination, we conducted repeated evaluations using 5 randomly selected seeds and computed the average accuracy. 
We chose accuracy as the evaluation metric in our experiments because the three datasets we used share a similar distribution of activity types and do not exhibit a long-tail category distribution. Thus, accuracy provides a reliable measure of overall model performance.

\subsection{Linear Separability of the Representations. (Address Q1)}
\label{sec:linear sepa}
To evaluate the quality of features obtained through SSL during the pre-training stage, we conducted experiments to analyze their linear separability. 
This provides insight into whether the learned features can be applied to various downstream tasks, rather than being limited to specific tasks directly related to the pre-training method. 
We investigated the effectiveness of separating the feature space acquired by SSL algorithms using a simple linear classifier, specifically a one-layer multi-layer perceptron. 
The underlying rationale is that a high-quality representation space learned through SSL naturally clusters samples of the same categories, obviating the need to fine-tune the representation space using label information. 
In such scenarios, a linear classifier alone should achieve satisfactory classification performance on the learned representation space.

To facilitate a comprehensive comparison of various SSL algorithms, we present the results in separate tables based on the specific encoder architecture utilized by each algorithm. It is important to highlight that Tables \ref{table:2} to \ref{table:5} showcase the highest accuracy achieved by an SSL algorithm in bold, while the second highest accuracy is indicated with an underline.

Table \ref{table:2} to \ref{table:4} display the results of algorithms using the Causal Net, ResNet and ViT, respectively.
We also present a summary of the optimal performance of all SSL algorithms on all encoder architectures in Tables \ref{table:2} to \ref{table:4} in Table \ref{table:5} for comparison. Based on the results presented in Table \ref{table:5}, it can be observed that SwAV achieves the highest accuracy of 87.50\% on the UT-HAR dataset, while MAE performs the best on both the SignFi and Widar\_R2 datasets, achieving accuracies of 88.77\% and 69.24\%, respectively. Additionally, MAE also achieves suboptimal performance on UT-HAR.

In summary, from the encoder architecture view, the overall trend is that ResNet generally outperforms Causal Net across the three contrastive algorithms. 
Furthermore, an examination of Table \ref{table:4} demonstrates that adopting VIT in the supervised baseline does not yield a significant overall improvement on the three datasets compared to using Causal Net and ResNet. 
However, a comparison between MoCo (ViT), MoCo (Causal), and MoCo (ResNet) reveals a substantial enhancement in performance on the SignFi and Widar\_R2 datasets when employing ViT as the encoder.
This implies that the ViT model are more suitable to SSL methods compared to supervised approach.
These observations show that it is advisable for practitioners to prioritize ResNet over Causal Net and similar networks.

From the perspective of downstream tasks, we find out that different methods perform differently on various tasks.
SwAV (ResNet) achieve the highest accuracy on daily activity recognition (UT-HAR). 
On the other hand, ViT-based models such as MAE and MoCo (ViT) show results that far outperform other SSL models on gesture recognition tasks (SignFi and Widar\_R2).
Therefore, researchers should consider SwAV (ResNet) as their preferred choice when dealing with daily activity detection tasks. However, when working with gesture recognition data, it is advisable to explore ViT-based SSL algorithms for potential improvements in performance.

Additionally, it is worth noting that the masked image modeling method MAE consistently outperforms the contrastive learning method MoCo when using the same encoder architecture. Furthermore, when compared to all other SSL algorithms that employ different encoder architectures, MAE consistently achieves either optimal or suboptimal results, highlighting its potential. 
These findings suggests researchers to pay more attention to the development of masked image modeling methods in WiFi-based HAR tasks.

\begin{table}[t]
    \label{Tab}
    \begin{subtable}{.45\linewidth}
        \centering
        \caption{}{
        \begin{tabular}{l|ccc}
\toprule
\multirow{2}{*}{Algorithm} & \multicolumn{3}{c}{Dataset}                                                         \\ 
                           & \multicolumn{1}{c}{UT-HAR} & \multicolumn{1}{c}{SignFi} & \multicolumn{1}{c}{Widar\_R2} \\ 
\midrule
Supervised                 & 98.03                      & 83.33                      & 85.48                      \\ 
SimCLR                     & 82.14                      & 46.45                      & 28.72                      \\ 
MoCo                       & \textbf{82.50}             & \textbf{62.64}             & \textbf{33.17}             \\ 
SwAV                       & 78.21                      & 58.41                      & 30.29                      \\ 
Rel-Pos                    & -                          & -                          & -                          \\ 
MAE                        & -                          & -                          & -                          \\ 
\bottomrule
\end{tabular}
        }
    \label{table:2}
    \end{subtable}

    \begin{subtable}{.45\linewidth}
        \centering
        \caption{}{
\begin{tabular}{l|ccc}
\toprule
\multirow{2}{*}{Algorithm} & \multicolumn{3}{c}{Dataset}                                                         \\ 
                           & \multicolumn{1}{c}{UT-HAR} & \multicolumn{1}{c}{SignFi} & \multicolumn{1}{c}{Widar\_R2} \\
\midrule
Supervised                 & 92.86                      & 97.65                      & 94.37                      \\
SimCLR                     & 83.93                      & 47.10                      & 40.43                      \\
MoCo                       & 82.14                      & 46.74                      & \textbf{41.13}             \\
SwAV                       & \textbf{87.50}             & 57.43                      & 36.39                      \\
Rel-Pos                    & 82.14                      & \textbf{81.34}             & 32.18                      \\
MAE                        & -                          & -                          & -                          \\
\bottomrule
\end{tabular}
        }
    \label{table:3}
    \end{subtable} 
    \medskip
        \begin{subtable}{.45\linewidth}
        \centering
        \caption{}{
    \begin{tabular}{l|ccc}
    \toprule
    \multirow{2}{*}{Algorithm} & \multicolumn{3}{c}{Dataset}                                                         \\ 
                               & \multicolumn{1}{l}{UT-HAR} & \multicolumn{1}{l}{SignFi} & \multicolumn{1}{l}{Widar\_R2} \\ 
    \midrule
    Supervised                 & 87.50                      & 89.31                      & 86.33                      \\ 
    SimCLR                     & -                          & -                          & -                          \\ 
    MoCo                       & 73.21                      & 87.14                      & 62.81                      \\ 
    SwAV                       & -                          & -                          & -                          \\ 
    Rel-Pos                    & -                          & -                          & -                          \\ 
    MAE                        & \textbf{84.29}             & \textbf{88.77}             & \textbf{69.24}             \\ 
    \bottomrule
    \end{tabular}
        }
    \label{table:4}
    \end{subtable} 
\caption{Linear separability of the learned representations using (a) Causal Net. (b) ResNet. (c) ViT. The best accuracy achieved by a self-supervised learning algorithm is denoted in bold. 
The "-" indicates that corresponding encoder architecture is not compatible with the algorithm specified in the respective row.
} 
\end{table}

\begin{center}
\begin{table}
\begin{tabular}{l|ccc}
\toprule
\multirow{2}{*}{Algorithm} & \multicolumn{3}{c}{Dataset}                                                         \\ 
                           & \multicolumn{1}{l}{UT-HAR} & \multicolumn{1}{l}{SignFi} & \multicolumn{1}{l}{Widar\_R2} \\ 
\midrule
Supervised                 & 98.03                      & 97.65                      & 86.33                      \\ 
SimCLR                     & 83.93                      & 47.10                      & 40.43                      \\ 
MoCo                       & 82.50                      & \underline{87.14}          & \underline{62.81}          \\ 
SwAV                       & \textbf{87.50}             & 58.41                      & 36.39                      \\ 
Rel-Pos                    & 82.14                      & 81.34                      & 32.18                      \\ 
MAE                        & \underline{84.29}          & \textbf{88.77}             & \textbf{69.24}       \\
\bottomrule
\end{tabular}
\caption{The optimal results of each algorithm, regardless of the encoder architecture, under the linear separability experiment. The best accuracy achieved by a self-supervised learning algorithm is denoted in bold, while suboptimal accuracy is marked with an underline.}
\label{table:5}
\end{table}
\end{center}


\subsection{Robustness to Domain Shift.  (Address Q2)}
WiFi-based HAR applications in smart homes, as discussed in Section \ref{sec:3.1.moti}, encounter the challenge of acquiring extensive labeled datasets. To address this, we propose leveraging the readily available unlabeled CSI dataset for pre-training, allowing the model to learn valuable weights, followed by fine-tuning on specific target application scenarios. This section aims to evaluate the performance of various SSL algorithms under this training approach, enabling a comprehensive comparison of their strengths and weaknesses in the given application context.

Our evaluation commences by assessing the influence of target user identity on recognition performance. Subsequently, we investigate the impact of variations in the room environment between the source and target datasets on activity recognition. Finally, we explore how discrepancies in the receiver position between the source and target datasets affect activity recognition accuracy.

All experiments in this section are conducted using a distinct subset of the Widar dataset, selected for its diversity. Both linear evaluation and fine-tuning results are provided for all experiments. In the context of linear evaluation, we keep the weights of the encoder fixed after pre-training, while in fine-tuning, the weights of the model are allowed to be updated.
The result of linear evaluation shows the 

Overall, these experiments shed light on the limitations of method transferability, offering valuable insights for practitioners in selecting the most suitable SSL algorithm for practical applications.

\subsubsection{User}
\label{sec:5.5.user}
In this section, our objective is to examine the efficacy of SSL algorithms in achieving satisfactory performance when there are variations in user identities between the unlabeled pre-training source dataset and the target dataset.

\begin{figure*}
     \centering
     \begin{subfigure}[b]{0.33\textwidth}
         \centering
         \includegraphics[width=\textwidth]{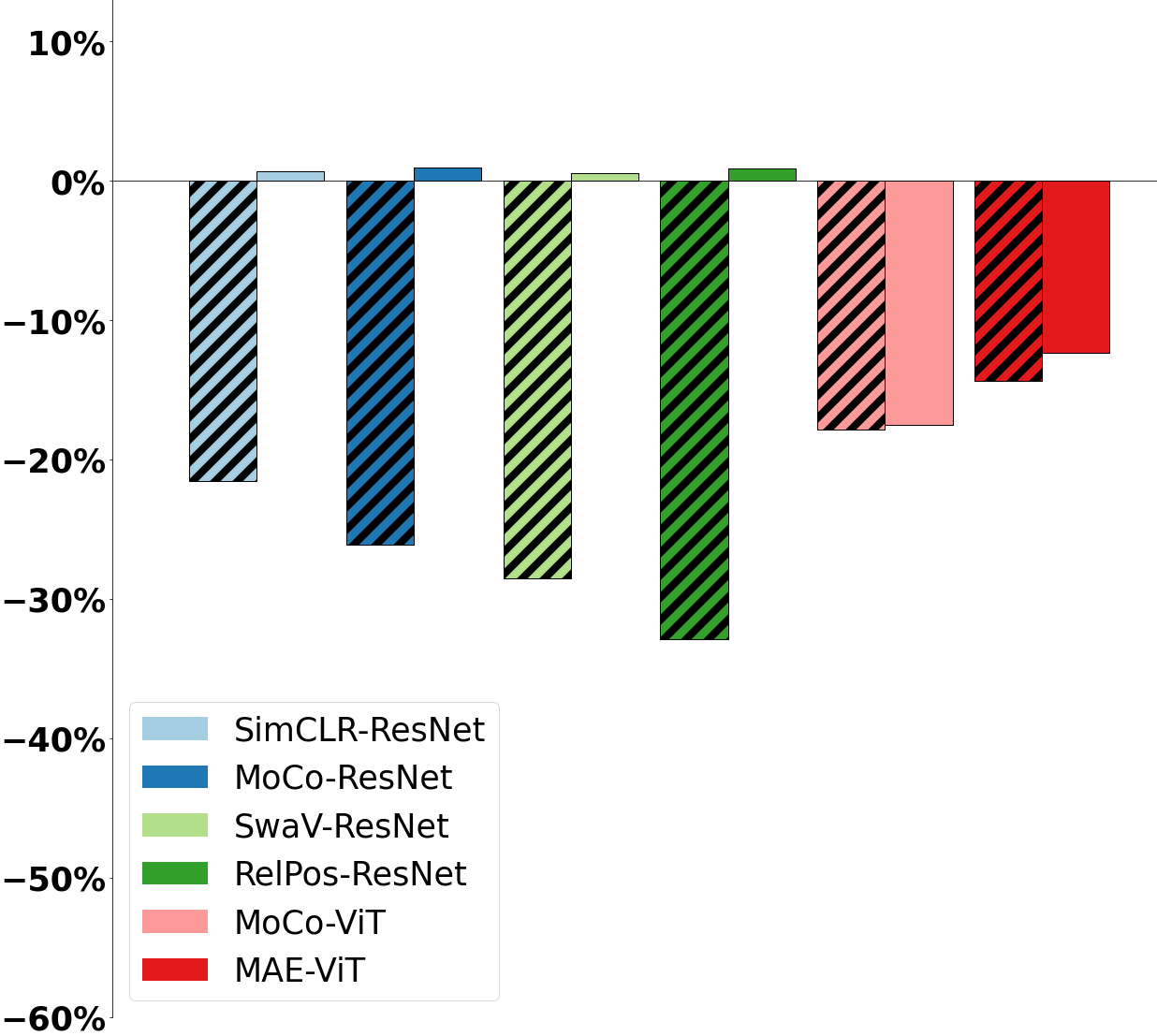}
         \caption{}
         \label{fig:user_a}
     \end{subfigure}
     \hfill
     \begin{subfigure}[b]{0.33\textwidth}
         \centering
         \includegraphics[width=\textwidth]{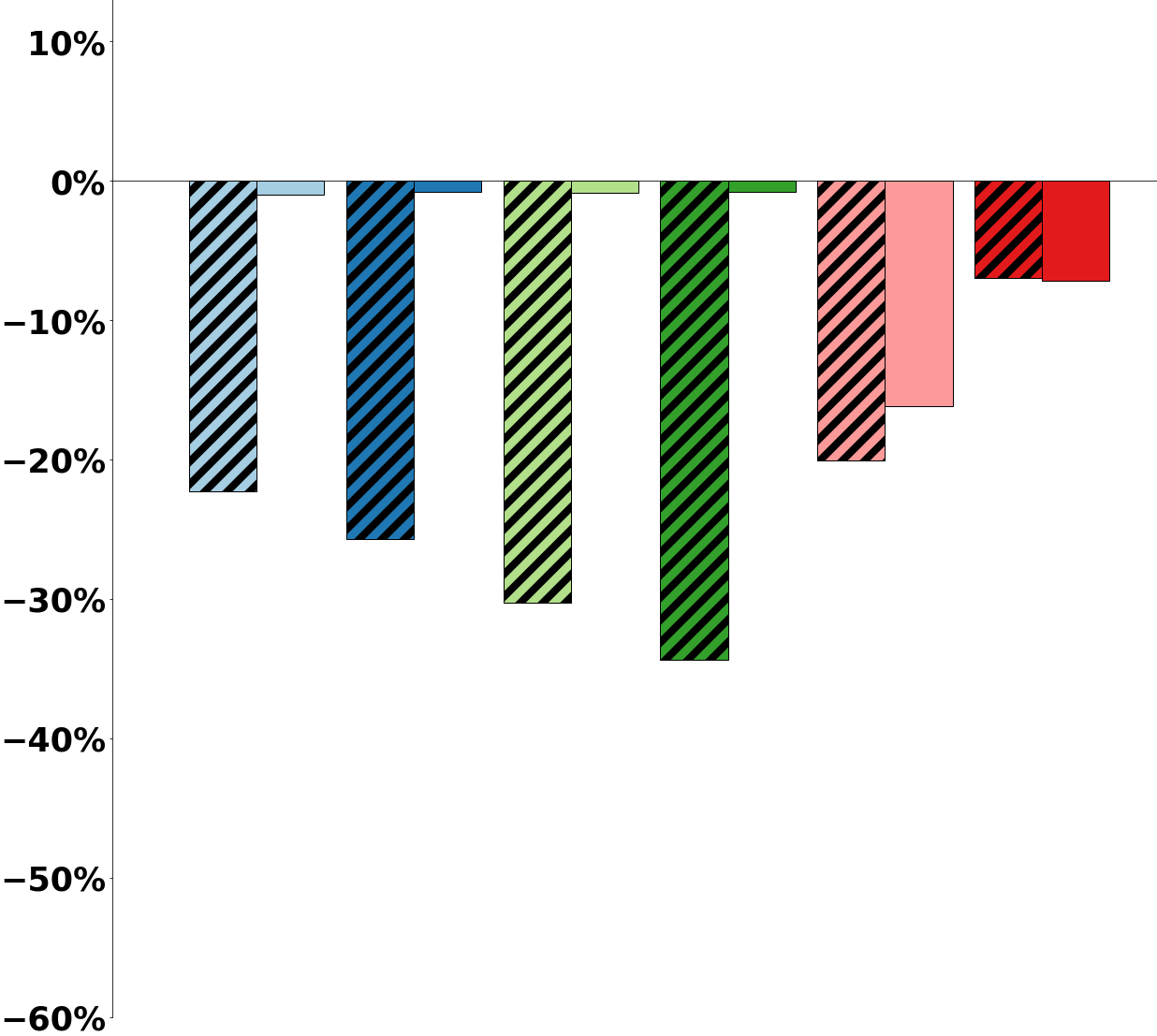}
         \caption{}
         \label{fig:user_b}
     \end{subfigure}
     \hfill
     \begin{subfigure}[b]{0.33\textwidth}
         \centering
         \includegraphics[width=\textwidth]{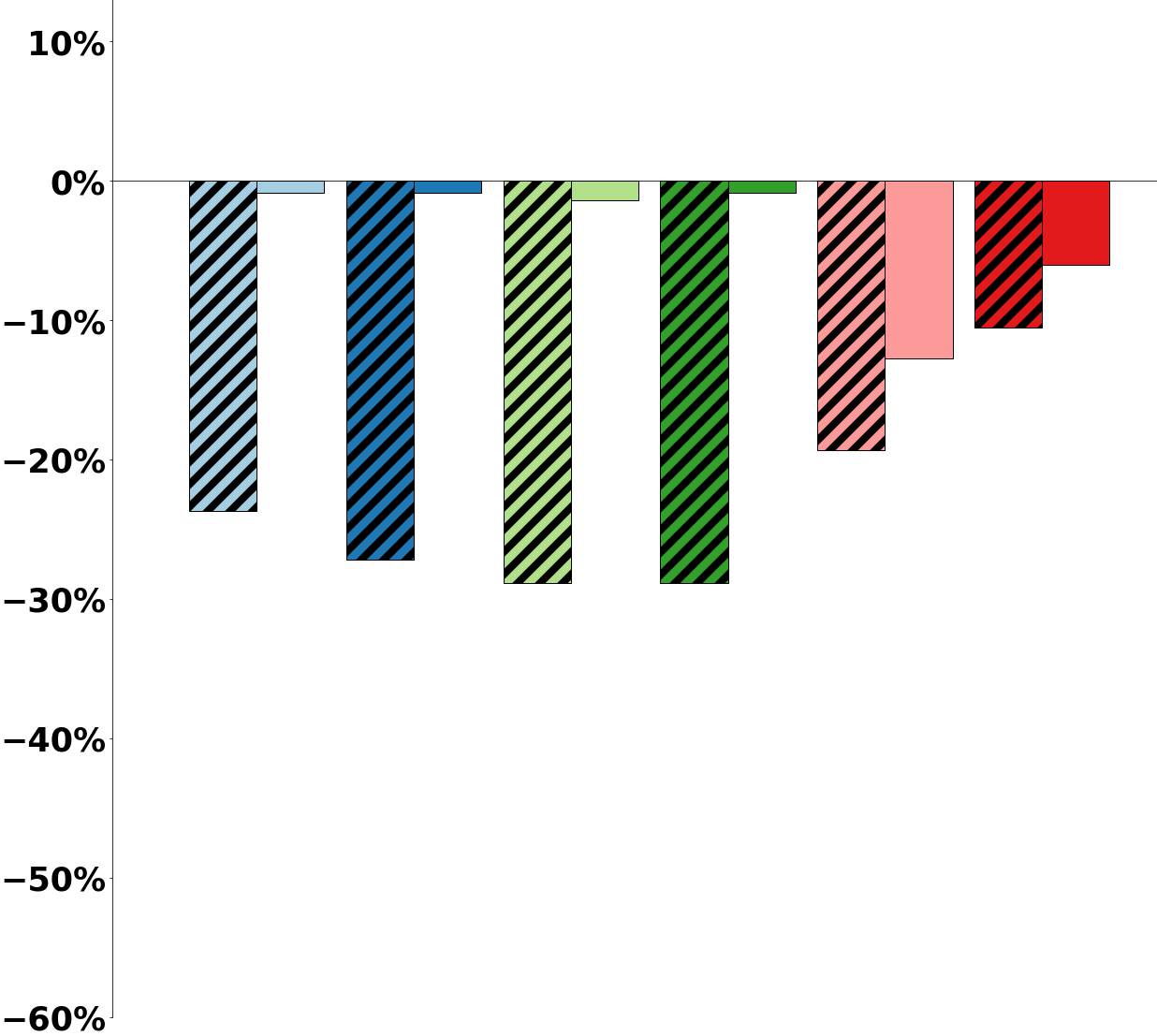}
         \caption{}
         \label{fig:user_c}
     \end{subfigure}
     \hfill
     \begin{subfigure}[b]{0.33\textwidth}
         \centering
         \includegraphics[width=\textwidth]{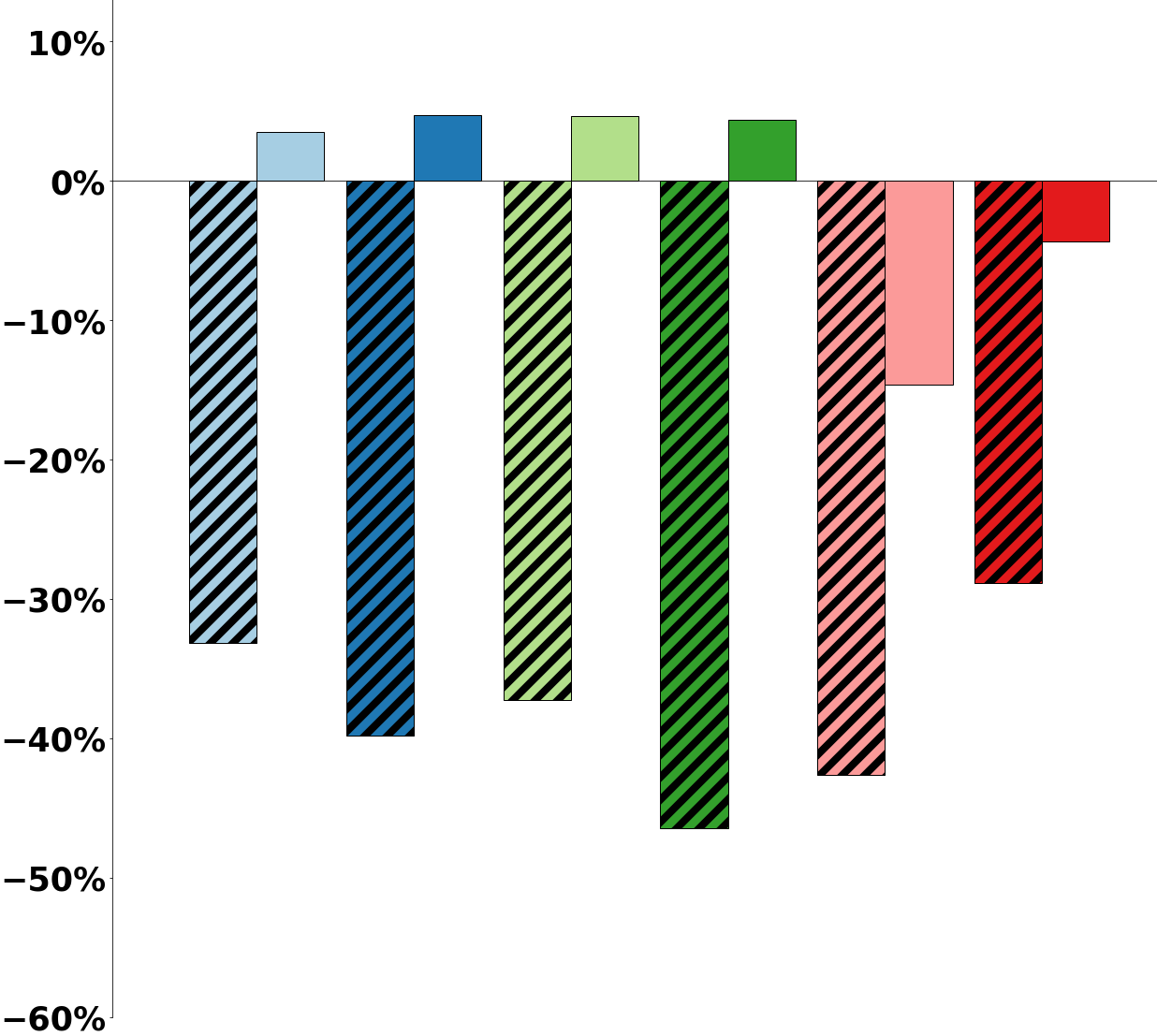}
         \caption{}
         \label{fig:user_d}
     \end{subfigure}
     \hfill
     \begin{subfigure}[b]{0.33\textwidth}
         \centering
         \includegraphics[width=\textwidth]{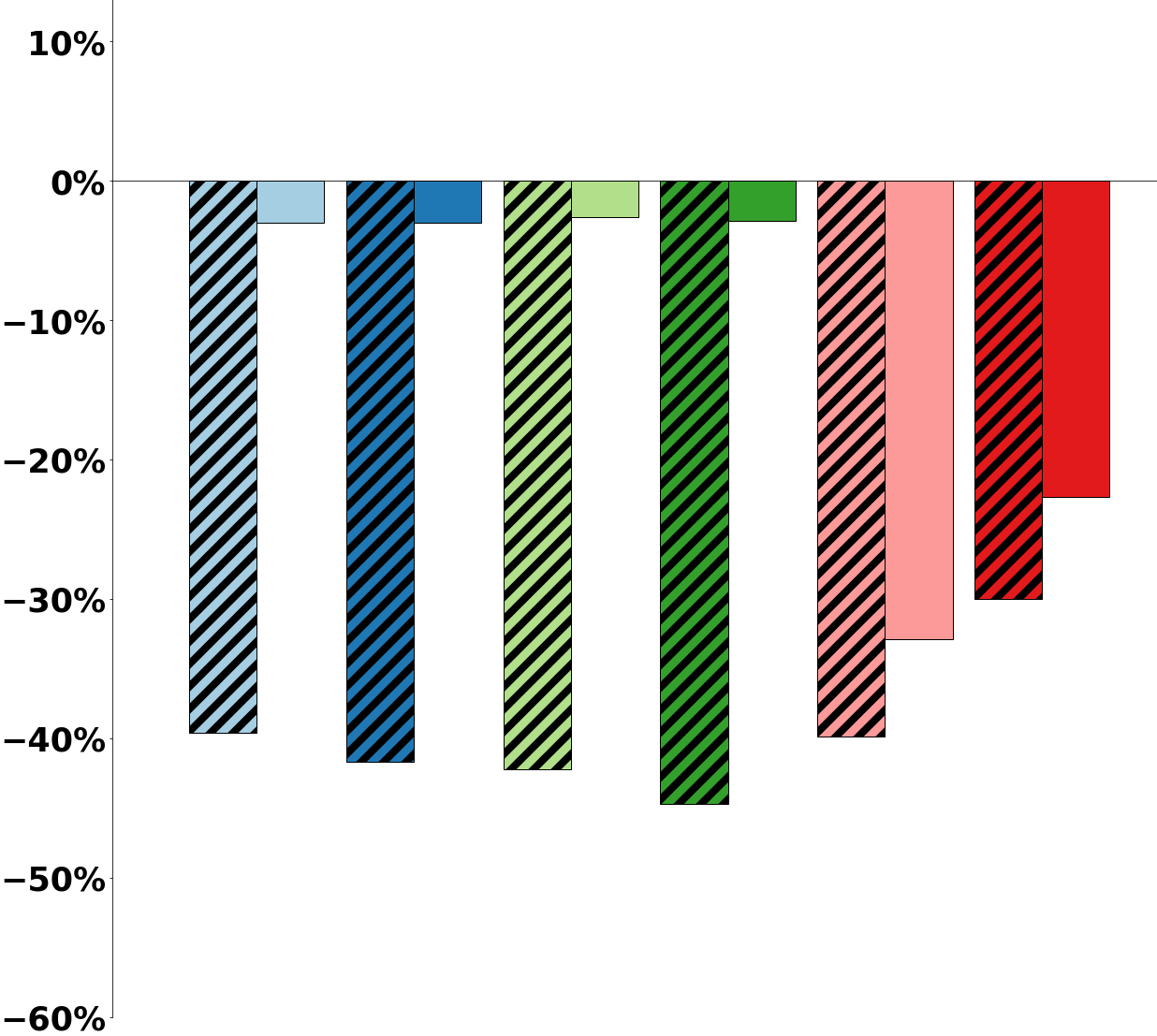}
         \caption{}
         \label{fig:user_e}
     \end{subfigure}
        \caption{Robustness of SSL algorithms to User change. The histogram presents the accuracy difference between the self-supervised learning algorithm and the supervised baseline using ResNet, that is $Accuracy_{SSL}-Accuracy_{Supervised}$. The bars filled with '/' indicates linear evaluation, while an unfilled bar represents the result of fine-tuning. Each subfigure presents a distinct transfer setting. (a) Transferring from Widar\_R1R2U1 to Widar\_R1R2U2. (b) Transferring from Widar\_R1R2U1 to Widar\_R1R2U3. (c) Transferring from Widar\_R1R2U2 to Widar\_R1R2U3. (d) Transferring from Widar\_R2R2U2 to Widar\_R2R2U1. (e) Transferring from Widar\_R2R2U2 to Widar\_R2R2U3.}
        \label{fig:User}
\end{figure*}

We controlled for variations in the room environment and receiver position within the Widar dataset, ensuring that the only distinction between the source dataset and the target dataset was the user identity.
Specifically, the SSL algorithms were pre-trained on source user A and then fine-tuned and tested on target user B.
For all experiments, the optimal hyperparameter combination and encoder architecture were selected for each SSL algorithm.
It is worth noting that we include the results of MoCo (ViT) as a control group, as it employs the same encoder architecture as MAE.
In Figure \ref{fig:User}, the accuracy difference between each SSL algorithm and the supervised baseline using ResNet ($Accuracy_{SSL}-Accuracy_{Supervised}$) is presented. The figure also includes histograms to differentiate the evaluation methods: a filled bar with '/' indicates linear evaluation, while an unfilled bar represents the result of fine-tuning.

The findings depicted in Figure \ref{fig:User} demonstrate  that the performance varies when transferring between different users. For example the performance gap in Figure \ref{fig:user_a},\ref{fig:user_b} and \ref{fig:user_c} is smaller than \ref{fig:user_d} and \ref{fig:user_e}.
Moreover, MAE has the smallest accuracy gap among all SSL algorithms in the linear evaluation setting. Moreover, when compared to the supervised baseline trained from scratch on the target user, MAE achieves very close accuracy in some cases. 
These results indicate that MAE produces a high-quality representation space, enabling robust performance even when there are changes in users for downstream tasks.

In the fine-tuning setting, Figure \ref{fig:User} illustrates that all algorithms utilizing the ResNet architecture outperformed the two algorithms employing ViT. 
These algorithms achieved accuracy levels that were very close to, and in some cases surpassed (i.e., achieving a positive accuracy gap), the supervised baseline. These observations suggest that the weights learned by SimCLR, MoCo, and SwAV during the SSL pre-training stage serve as excellent initial weights and contribute to the acquisition of superior model weights during the fine-tuning stage. 
However, our findings indicate that the performance improvement of ViT's fine-tuned model is often not as significant as that of the ResNet-based algorithms, compared to the linear evaluation that solely trains classifiers. We speculate that this may be attributed to the larger model capacity of ViT compared to ResNet and the insufficient amount of data available for fine-tuning, resulting in reduced performance when transferring to new users or environments. 
Based on these experimental results, we recommend using ResNet-based models for practical applications when the amount of data for the target user is insufficient to support fine-tuning of ViT. This study's findings not only demonstrate the potential of ViT models but also highlight their limitations in terms of data efficiency, thereby suggesting a new research direction for the future.

%

\subsubsection{Room}
To what extent do SSL algorithms effectively achieve satisfactory performance when the room environment in the unlabeled pre-training source dataset differs from the target dataset in a specific deployment scenario?

\begin{figure*}
     \centering
     \begin{subfigure}[b]{0.33\textwidth}
         \centering
         \includegraphics[width=\textwidth]{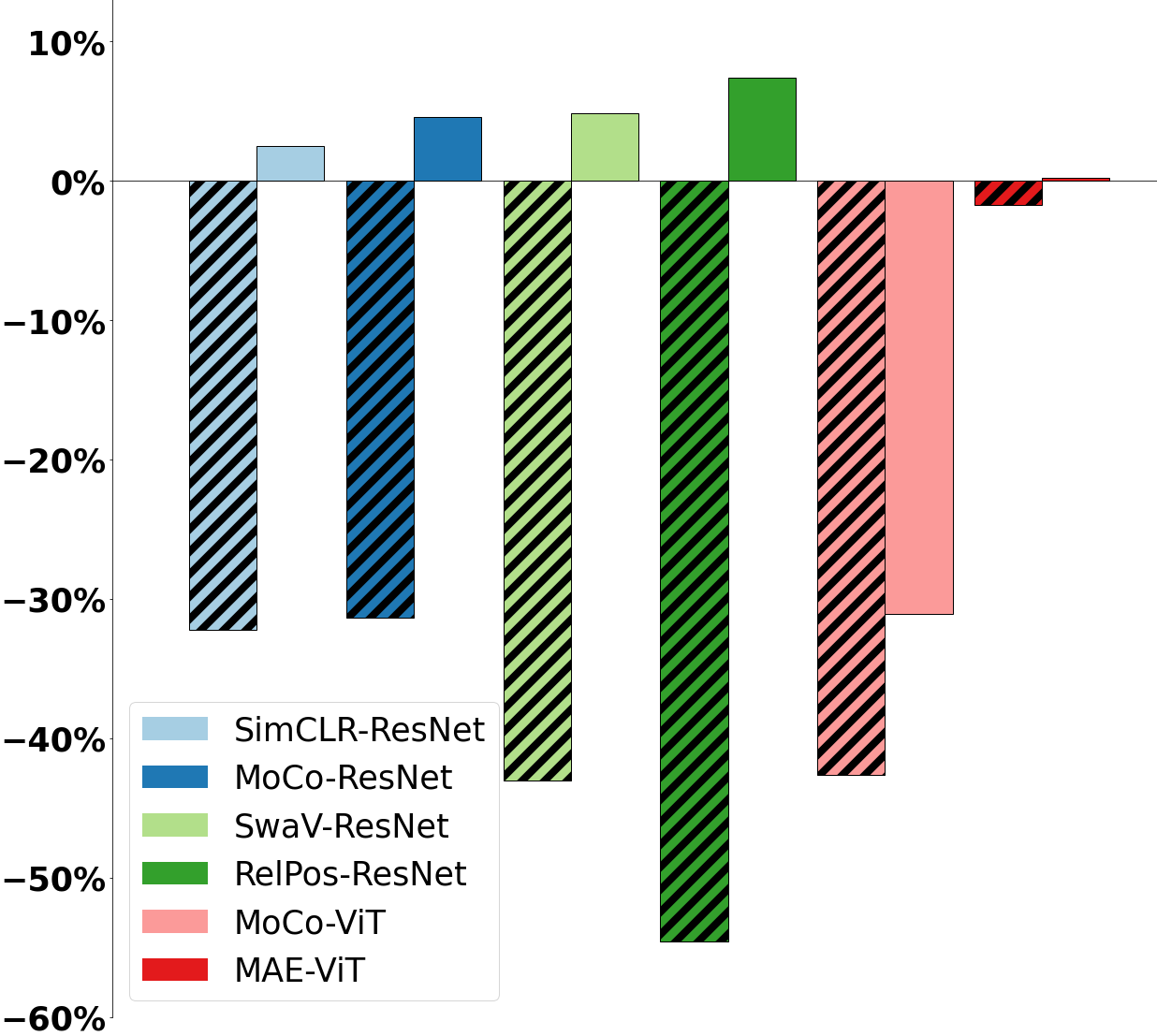}
         \caption{}
         \label{fig:room_a}
     \end{subfigure}
     \hfill
     \begin{subfigure}[b]{0.33\textwidth}
         \centering
         \includegraphics[width=\textwidth]{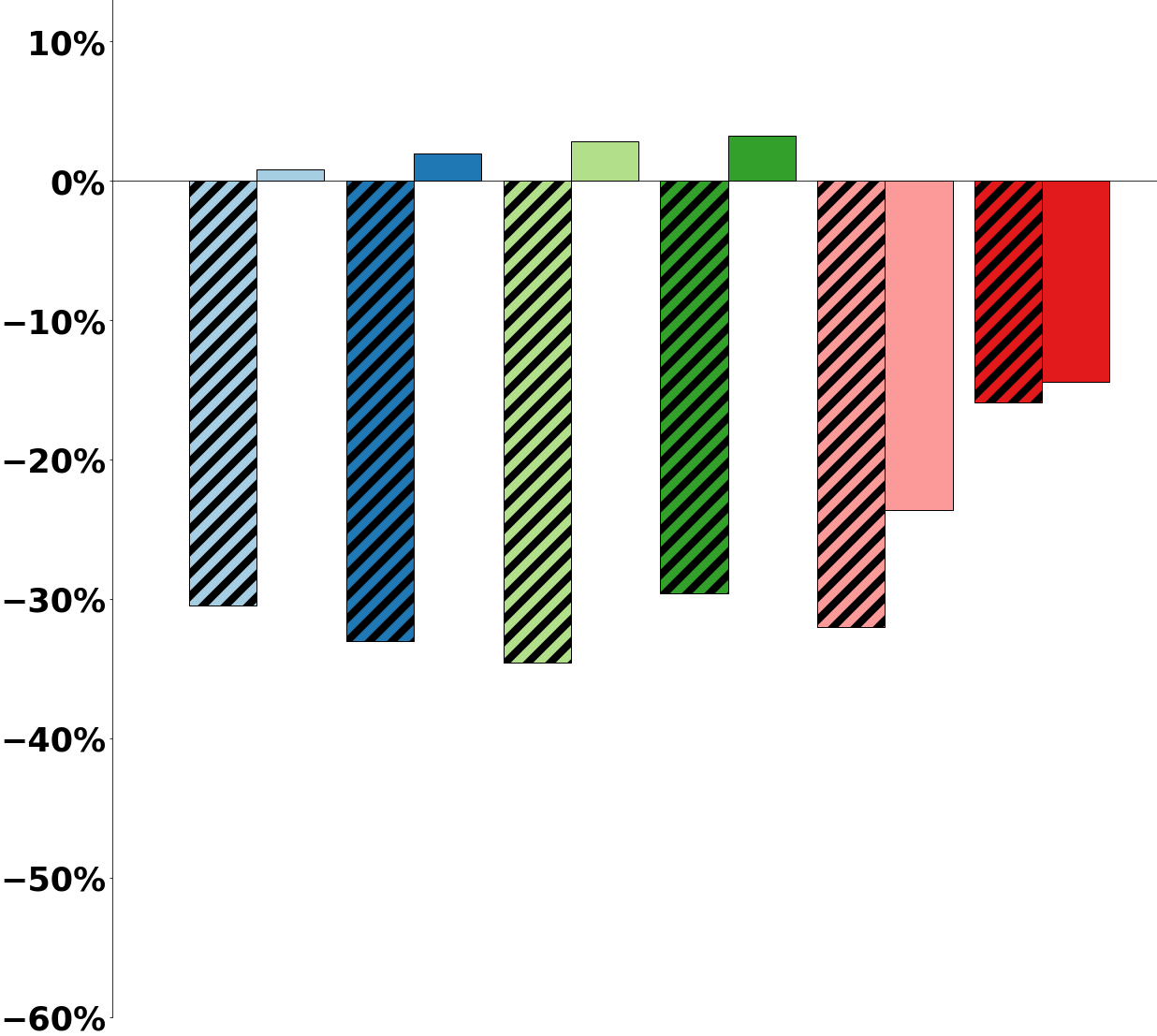}
         \caption{}
         \label{fig:room_b}
     \end{subfigure}
     \hfill
     \begin{subfigure}[b]{0.33\textwidth}
         \centering
         \includegraphics[width=\textwidth]{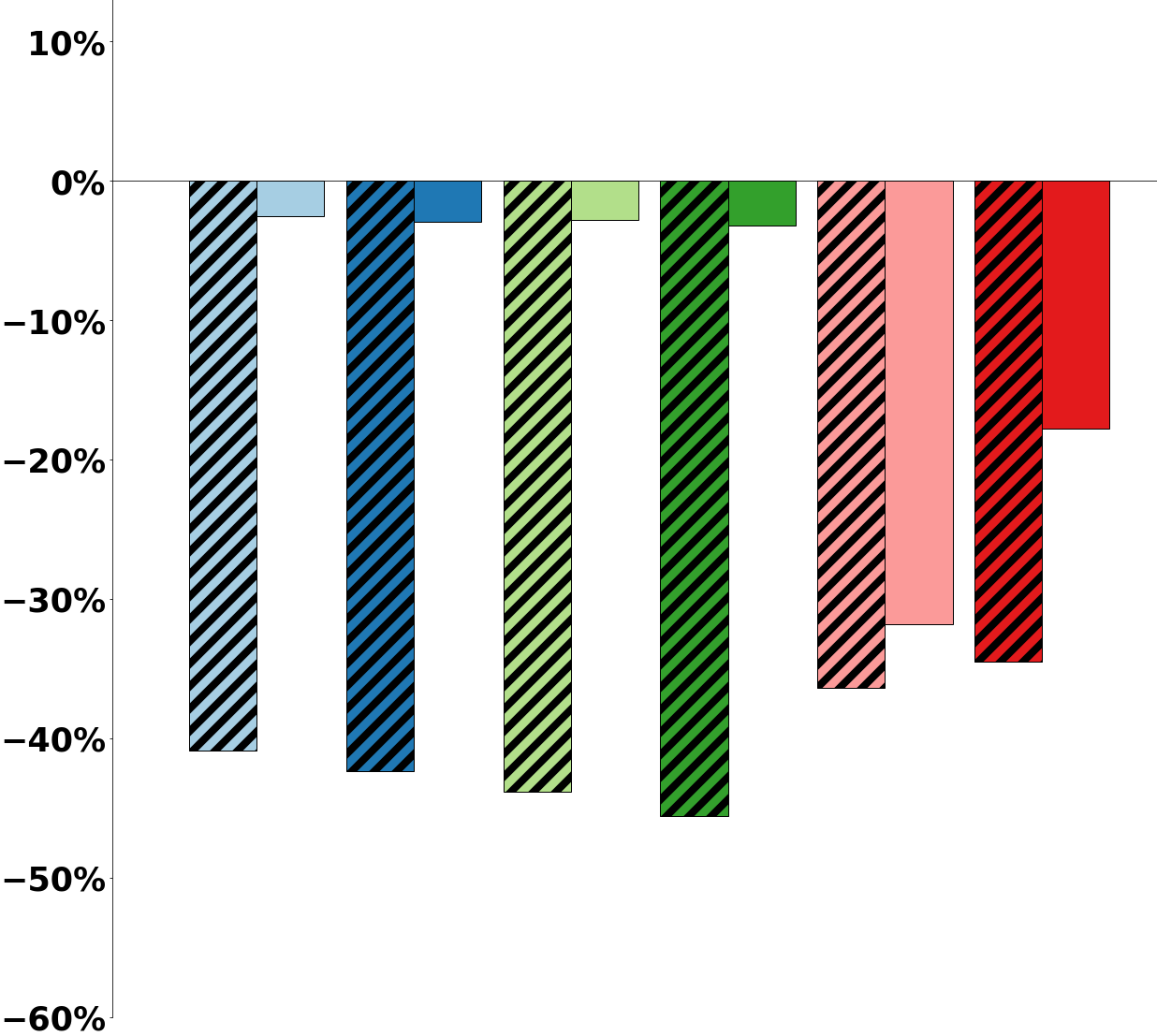}
         \caption{}
         \label{fig:room_c}
     \end{subfigure}
     \hfill
     \begin{subfigure}[b]{0.33\textwidth}
         \centering
         \includegraphics[width=\textwidth]{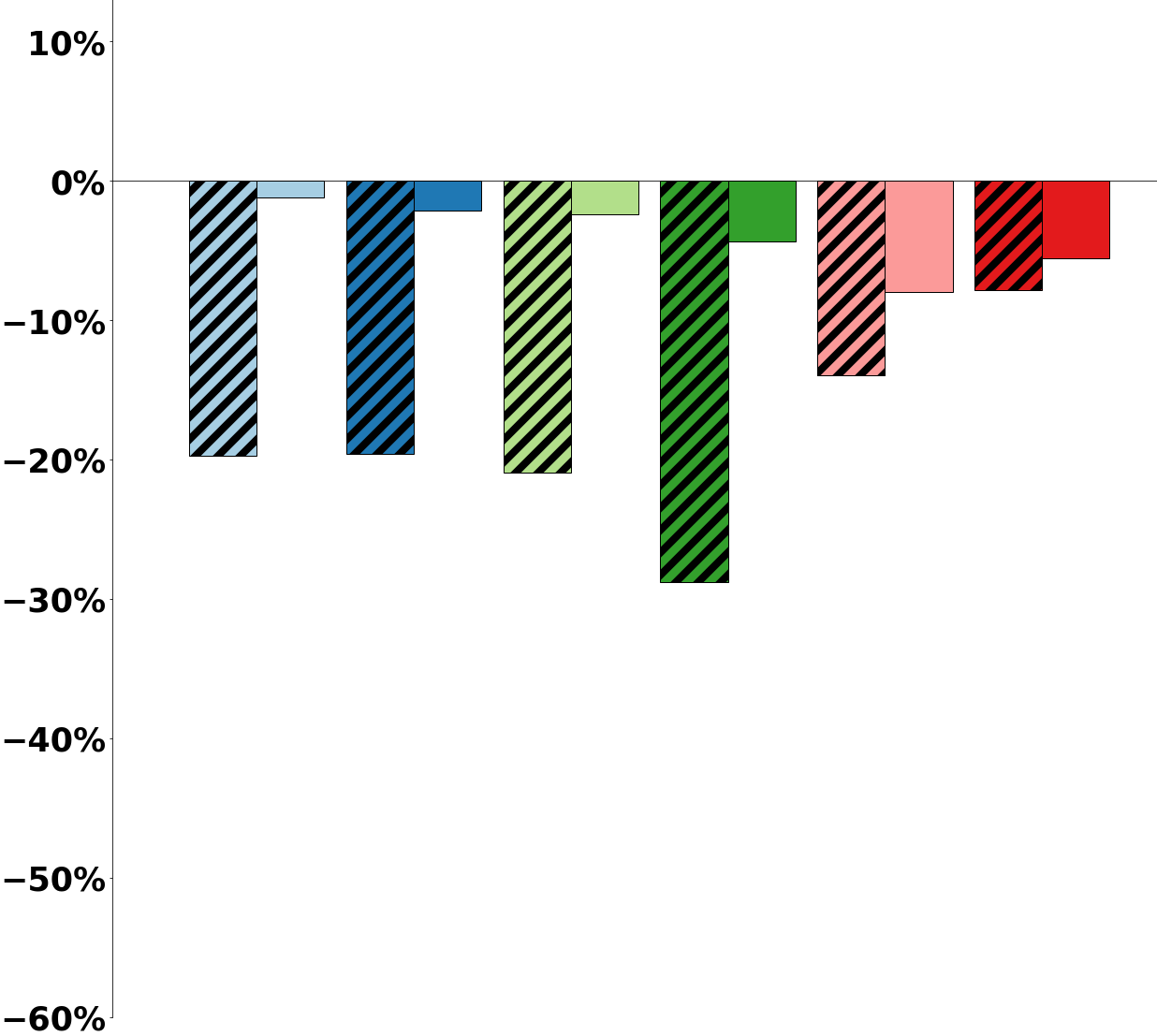}
         \caption{}
         \label{fig:room_d}
     \end{subfigure}
     \hfill
     \begin{subfigure}[b]{0.33\textwidth}
         \centering
         \includegraphics[width=\textwidth]{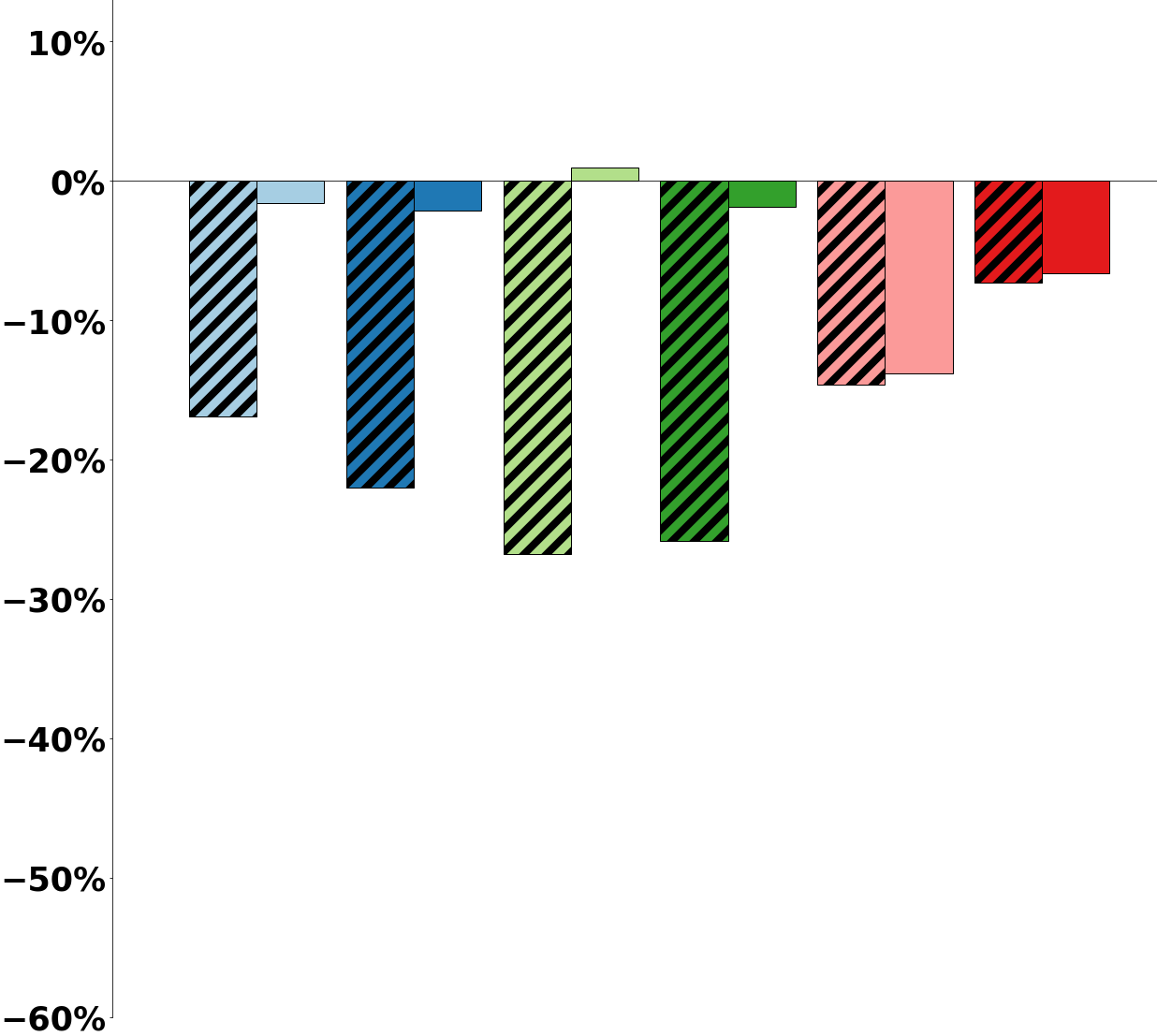}
         \caption{}
         \label{fig:room_e}
     \end{subfigure}
        \caption{Robustness of SSL algorithms to Room change. The histogram presents the accuracy difference between the self-supervised learning algorithm and the supervised baseline using ResNet. The bars filled with '/' indicates linear evaluation, while an unfilled bar represents the result of fine-tuning. Each subfigure presents a distinct transfer setting. 
        (a) Transferring from Widar\_R1R2U1 to Widar\_R2R2U1. 
        (b) Transferring from Widar\_R1R2U2 to Widar\_R2R2U2.
        (c) Transferring from Widar\_R1R2U3 to Widar\_R2R2U3.
        (d) Transferring from Widar\_R1R2U3 to Widar\_R3R2U3.
        (e) Transferring from Widar\_R2R2U3 to Widar\_R3R2U3.}
        \label{fig:Room}
\end{figure*}

To investigate this question, we conducted a series of experiments while controlling for variations in user identity and receiver position in the Widar dataset. This ensured that the sole distinction between the source and target datasets was the room environment. The experimental results of the SSL algorithms can be found in Figure \ref{fig:Room}. 

Figure \ref{fig:Room} presents the linear evaluation results. It shows the variability in performance when transferring between different rooms.
The superior performance of MAE across all room environment transfer settings are also demonstrated in Figure \ref{fig:Room}. 
Notably, MAE demonstrates a substantial advantage over other methods in specific cases, as illustrated in Figure \ref{fig:room_a}, where it outperforms other methods by over $29\%$. 

The results in Figure \ref{fig:Room} indicate that in the fine-tuning setting, SSL algorithms leveraging the ResNet consistently outperform those utilizing ViT. The top-performing algorithms achieve results comparable to the supervised baseline, which exhibits an accuracy gap close to 0 or even positive values. 
Comparing these results with the linear evaluation outcomes reveals that models utilizing the ResNet architecture exhibit an accuracy improvement of over $17\%$, whereas the improvement in models using ViT is less pronounced.


\subsubsection{Receiver}
In this section, our objective is to examine the impact of receiver position shifts on model performance.

\begin{figure*}
     \centering
     \begin{subfigure}[b]{0.45\textwidth}
         \centering
         \includegraphics[width=\textwidth]{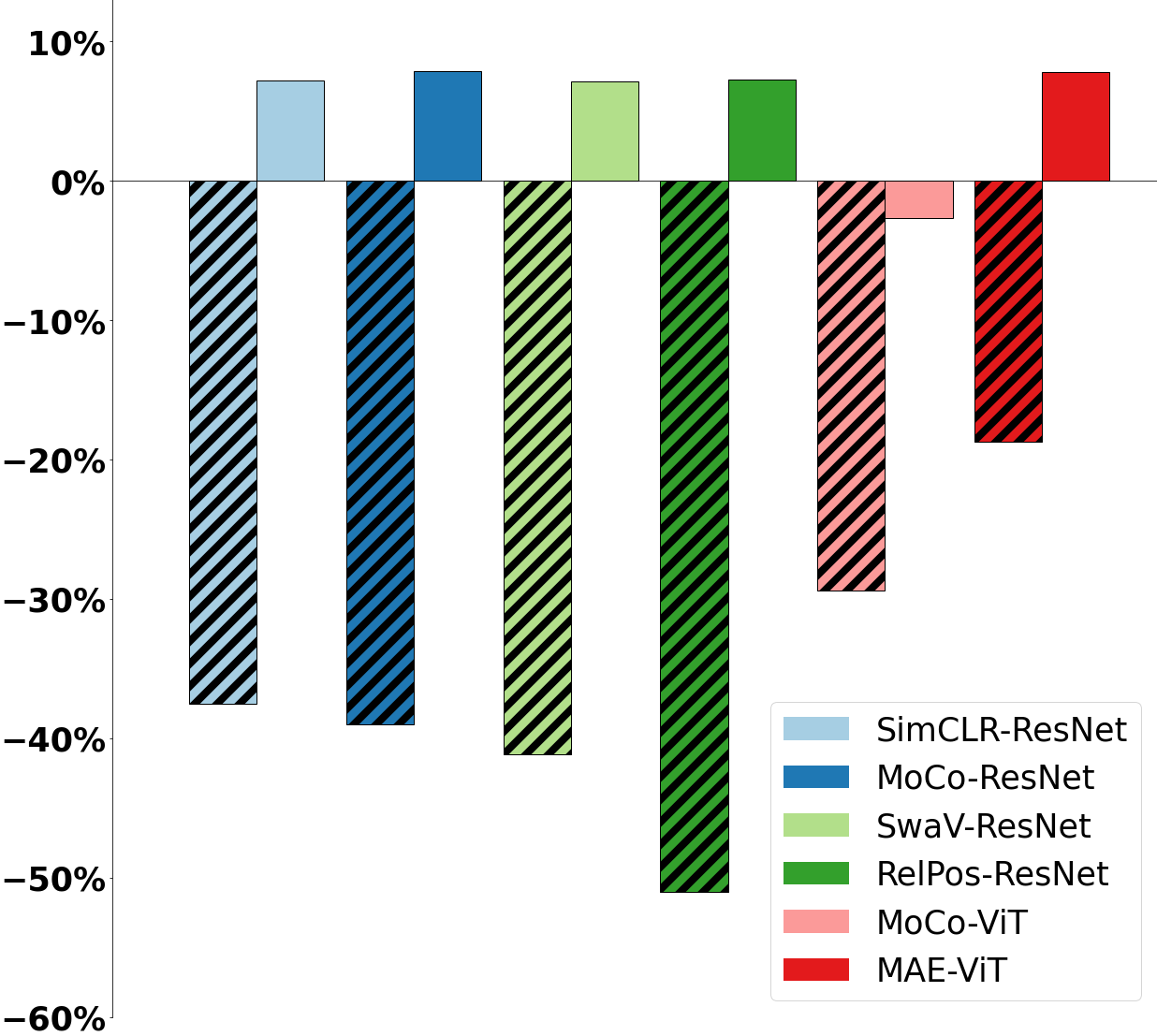}
         \caption{}
         \label{fig:receiver_a}
     \end{subfigure}
     \hfill
     \begin{subfigure}[b]{0.45\textwidth}
         \centering
         \includegraphics[width=\textwidth]{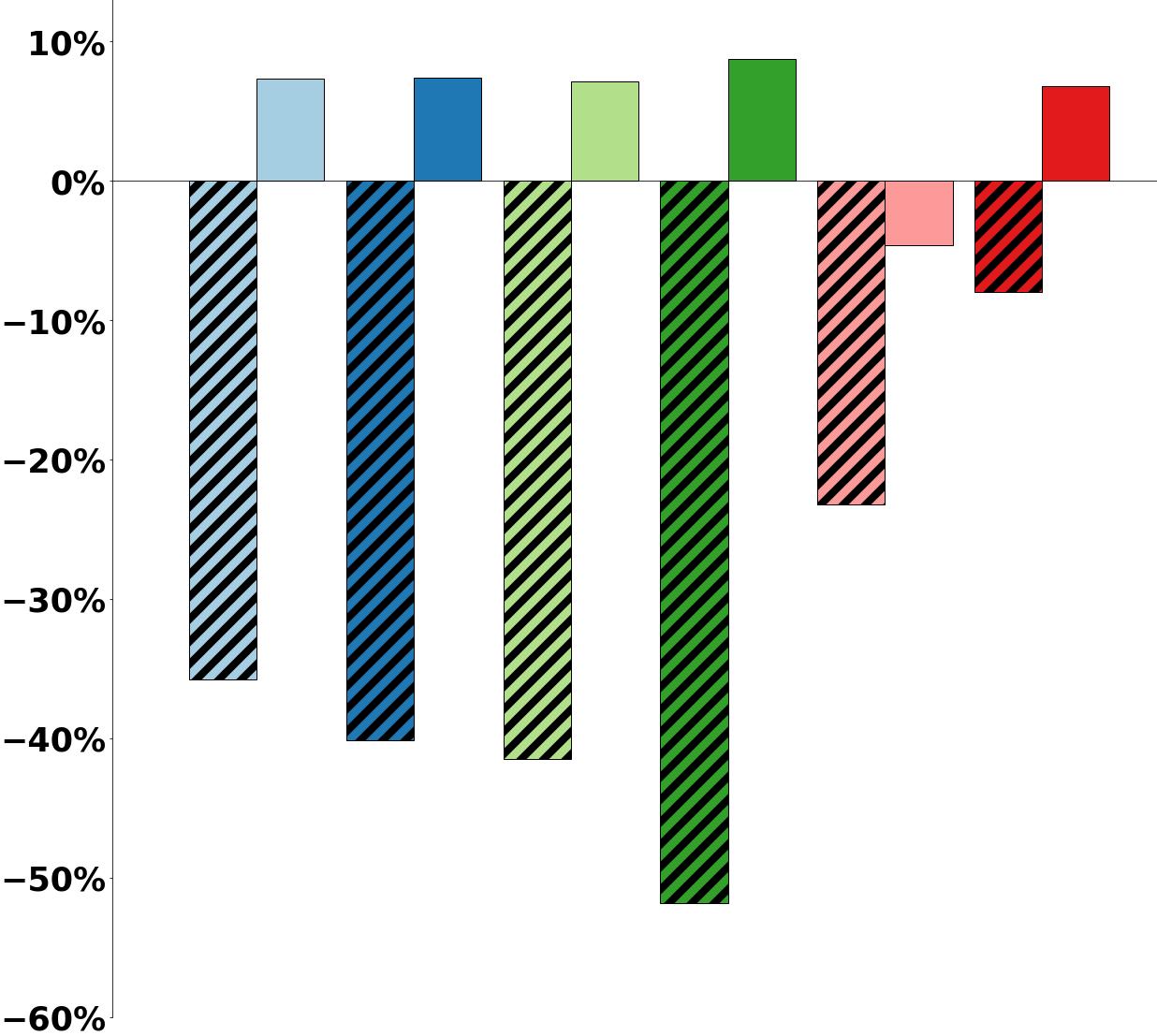}
         \caption{}
         \label{fig:receiver_b}
     \end{subfigure}
     \hfill
     \begin{subfigure}[b]{0.45\textwidth}
         \centering
         \includegraphics[width=\textwidth]{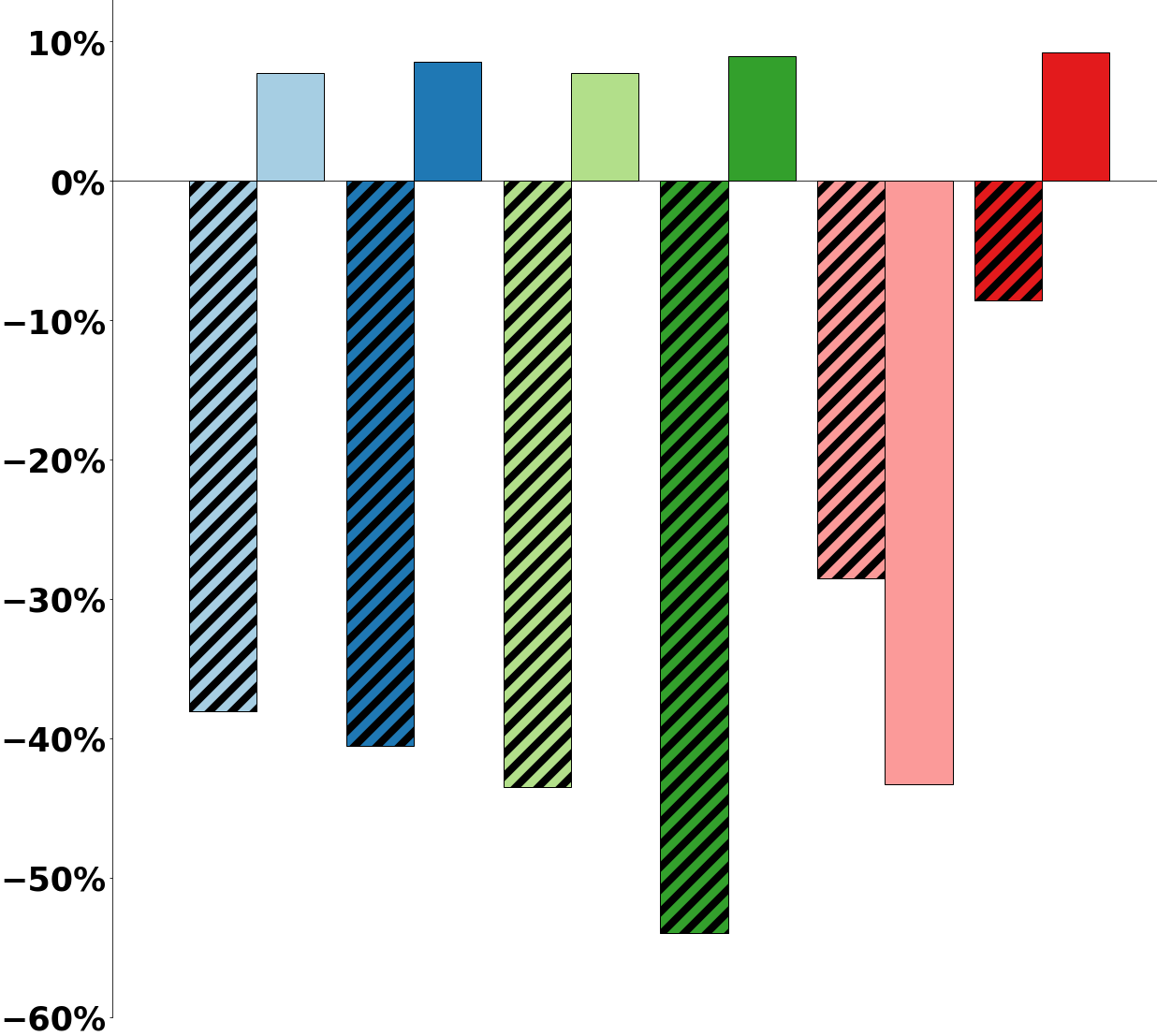}
         \caption{}
         \label{fig:receiver_c}
     \end{subfigure}
     \hfill
     \begin{subfigure}[b]{0.45\textwidth}
         \centering
         \includegraphics[width=\textwidth]{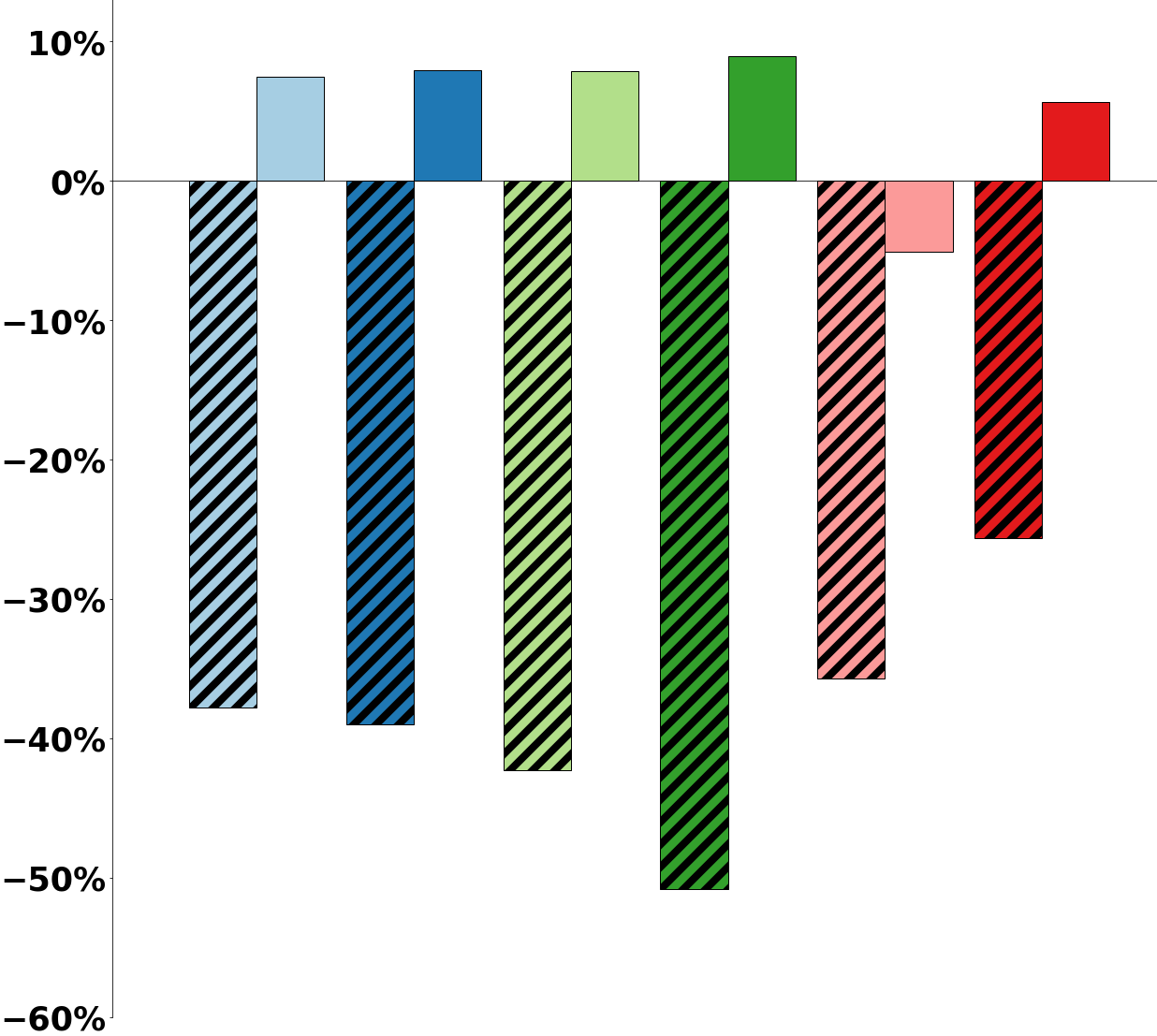}
         \caption{}
         \label{fig:receiver_d}
     \end{subfigure}
     \caption{Robustness of SSL algorithms to Receiver change. The histogram presents the accuracy difference between the self-supervised learning algorithm and the supervised baseline using ResNet. The bars filled with '/' indicates linear evaluation, while an unfilled bar represents the result of fine-tuning. Each subfigure presents a distinct transfer setting. 
        (a) Transferring from Widar\_R1R1 to Widar\_R1R2.
        (b) Transferring from Widar\_R1R2 to Widar\_R1R1.
        (c) Transferring from Widar\_R1R2 to Widar\_R1R5.
        (d) Transferring from Widar\_R1R5 to Widar\_R1R2.}
     \label{fig:Receiver}
\end{figure*}

Among the six available receiver positions in the Widar dataset, we specifically chose positions {1, 2, 5} situated in Room 1 and encompassing data from all participants.
Figure \ref{fig:Receiver} displays the experimental results of SSL algorithms.

In linear evaluation settings of Figure \ref{fig:Receiver}, RelPos exhibits particularly poor results.
SimCLR, MoCo (ResNet), and SwAV show diminishing but similar performance. MAE once again achieves the best performance on all target datasets under the linear evaluation setting, followed by MoCo (ViT).

Based on the observations in the fine-tuning setting, it is evident that all SSL algorithms, with the exception of MoCo (ViT), consistently outperform the supervised baseline that uses ResNet. This finding confirms the effectiveness of SSL pre-training in providing valuable information and generating superior initial weights for subsequent fine-tuning.
In the majority of fine-tuning cases in this section, SSL outperforms the supervised baseline by more than 7\%. 
Furthermore, contrary to the results obtained in previous sections, MAE not only achieves the best performance in the linear evaluation setting but also demonstrates competitive performance after fine-tuning in certain cases (e.g., in Figure \ref{fig:receiver_a} and \ref{fig:receiver_c}).
Although the accuracy improvement between the linear evaluation and fine-tuning stages of MAE is still not as pronounced as that observed in algorithms utilizing ResNet, it is worth noting.
We speculate that the excellent fine-tuning performance of MAE in such transfer settings is attributable to the availability of sufficient target data (over 30 thousand records, as shown in Table \ref{table:1}), enabling ViT to be well-trained during the fine-tuning stage.


\subsubsection{Section Summary}
Overall, we can summarize and analyze the experiments conducted in this section as follows.

Firstly, in the linear evaluation phase, MAE consistently achieves the best performance across all transfer settings, including user, room, and receiver position. This result indicates the high quality and robustness of the representation space learned by MAE. Given the performance exhibited in various transfer settings, researchers should prioritize exploring the potential of MAE and devote more attention to harnessing its capabilities in this area.

Secondly, due to the inherent complexity of the ViT model, MAE may exhibit suboptimal performance in fine-tuning when the target dataset is limited. However, when an ample amount of target data is available, MAE has the potential to achieve the highest level of performance after fine-tuning. Therefore, when dealing with a small target dataset, ResNet-based methods should be the primary choice for practitioners.

Thirdly, in most cases, SSL algorithms can reach the same level or surpass the supervised baseline with the ResNet. This finding suggests that SSL algorithms can enhance model performance by leveraging the information in the unlabeled source dataset.

Finally, in all transfer settings, when utilizing ViT as encoder, MoCo (ViT) has the better linear evaluation performance and worse fine-tuning performance compared with MoCo (ResNet). Meanwhile, compared to MAE, MoCo (ViT) underperforms in all situation.
These two observations indicate that the ViT model is capable of capturing the intrinsic characteristics of CSI data more effectively in the context of self-supervision and can learn to represent the space more accurately. 
In comparison to instance discrimination (MoCo), the Autoencoder algorithm (i.e., MAE) demonstrates a higher level of unsupervised learning ability, which suggests that the existing augmentation function may not be well-suited for CSI data.

\subsection{Robustness of SSL Algorithms to Data Scarcity. (Address Q3)}

\subsubsection{Robustness to Limited Labeled Data}
\label{sec:4.4.1.limited label data}
Undoubtedly, collecting labeled CSI data poses significant challenges. Therefore, this section focuses on examining the robustness of SSL when the availability of labeled target data decreases. Our objective is to determine which SSL algorithm demonstrates the highest efficiency in utilizing the limited labeled target data.

\begin{figure*}
     \centering
     \begin{subfigure}[b]{0.49\textwidth}
         \centering
         \includegraphics[width=\textwidth]{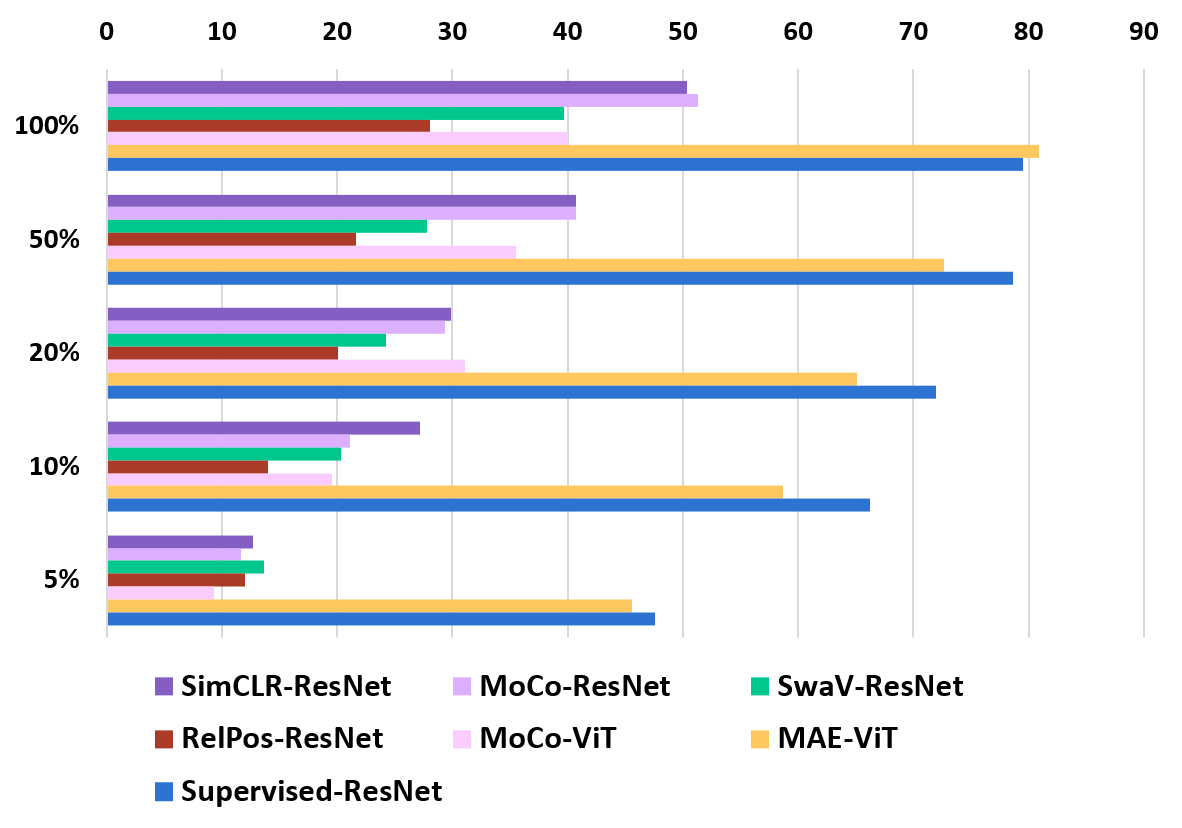}
         \caption{The accuracy of the linear evaluation setting with varying target training dataset size.}
         \label{fig:labeled_linear_a}
     \end{subfigure}
     \hfill
     \begin{subfigure}[b]{0.49\textwidth}
         \centering
         \includegraphics[width=\textwidth]{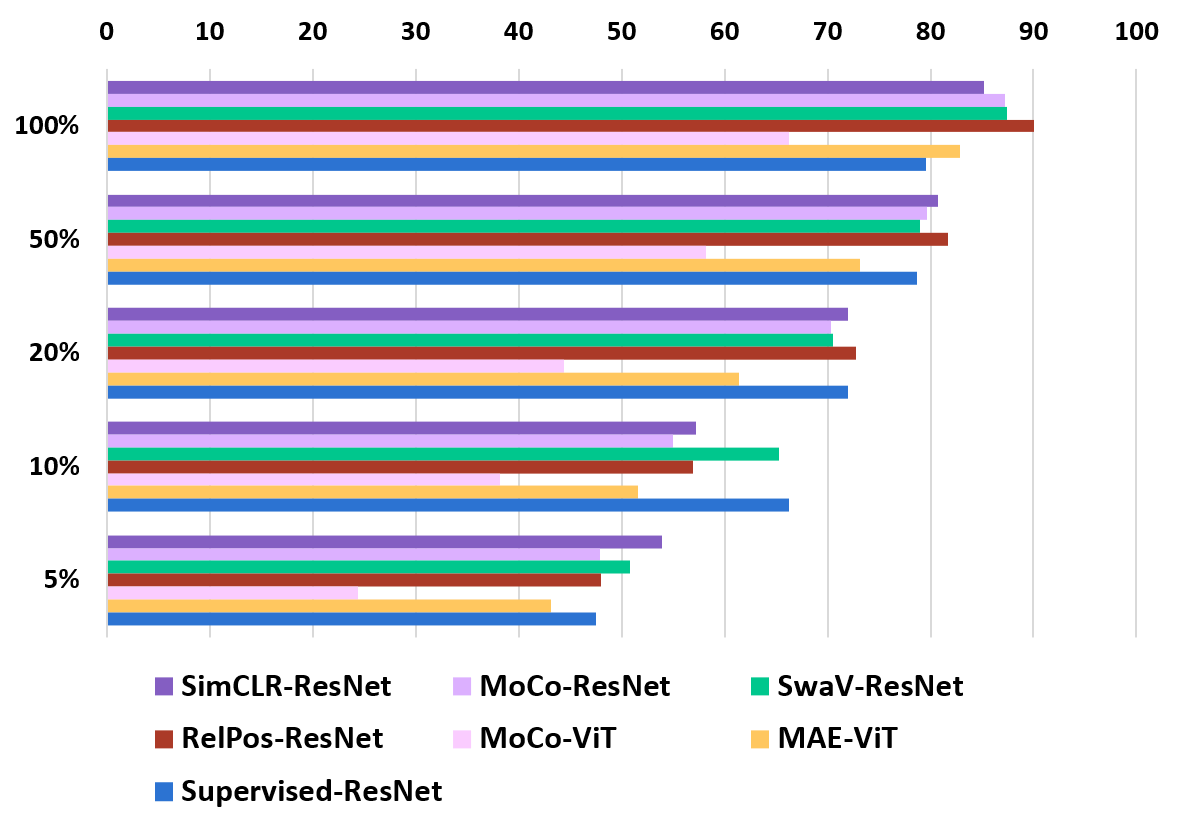}
         \caption{The accuracy of the fine-tuning setting with varying target training dataset size.}
         \label{fig:labeled_ft_b}
     \end{subfigure}
     \vfill
     \begin{subfigure}[b]{\textwidth}
         \centering
         \includegraphics[width=\textwidth]{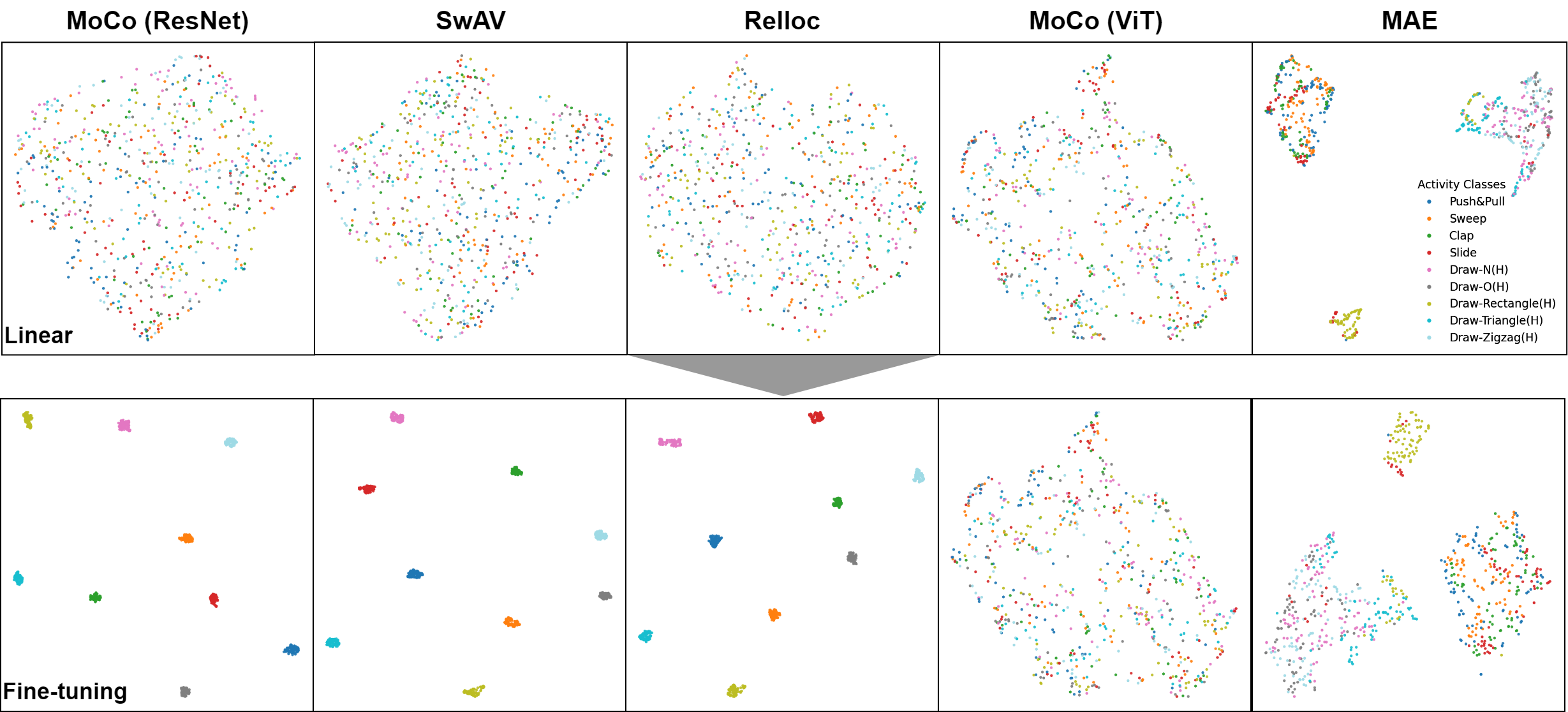}
         \caption{Learned representation under 100\% target training dataset size of linear evaluation and fine-tuning setting.}
         \label{fig:tsne}
     \end{subfigure}
        \caption{Analysis of Labeled Data Efficiency. Quantitative accuracy with varying sizes of the target training dataset. The model was trained on the source dataset collected from User 1 on Receiver 2 in Room 1, while the target dataset was collected in Room 2. (a) and (b) depict the results of linear evaluation and fine-tuning, respectively. (c) compares the learned representation between linear evaluation and fine-tuning using a 100\% target training dataset.}
        \label{fig:limitedlabeled}
\end{figure*}

In our experiment, we used Widar\_R1R2U1 as the source unlabeled dataset and Widar\_R2R2U1 as the target dataset. 
Both datasets consist of data collected from User 1 on Receiver 2. 
To assess the impact of limited labeled data, we gradually reduced the size of the target training dataset from 100\% to 5\%. 
The bar charts in Figure \ref{fig:limitedlabeled} display the performance results of all SSL algorithms and the ResNet supervised baseline. 
The supervised baseline results shown in Figure \ref{fig:labeled_linear_a} and \ref{fig:labeled_ft_b} are identical, indicating training from scratch using ResNet. 
We selected ResNet for the supervised baseline due to its superior performance.
Figure \ref{fig:tsne} depicts the UMAP visualization of the feature space for SSL algorithms of the linear evaluation and the fine-tuning setting, utilizing 100\% labeled data.

Upon analyzing Figure \ref{fig:labeled_linear_a}, it is evident that MAE achieves performance that is only slightly inferior to the supervised baseline. 
Distance discrimination algorithms, namely SimCLR and MoCo, are considered suboptimal SSL methods. However, their performance deteriorates and reaches a similar level as cluster discrimination and relation prediction methods when the target training dataset size is small.

Moreover, Figure \ref{fig:labeled_ft_b} demonstrates that SSL, specifically instance discrimination and relation prediction, outperforms the supervised baseline in the fine-tuning setting, except when the data size is 10\%. In contrast, MAE, which excels in linear evaluation, only surpasses the supervised baseline when the data size is 100\%.

Upon observing the UMAP analysis in Figure \ref{fig:tsne}, it becomes evident that all algorithms, with the exception of MAE, exhibit poor feature space quality during linear evaluation. Conversely, MAE demonstrates pronounced clustering within the same category in its feature space.
After fine-tuning, MoCo (ResNet), SwAV, and Relloc demonstrate significant improvement in their feature spaces, while the feature space of MAE remains largely unchanged. Notably, MoCo (ViT) fails to achieve a better feature space representation effect in both linear evaluation and fine-tuning.

\subsubsection{Robustness to Limited Unlablled Data}
In real-world application environments, an essential factor to consider is the accessbility and cost of 
larget scale unlabeled WiFi CSI dataset.
Compared to the computer vision field, where online platforms such as Instagram can be leveraged to collect large amounts of unlabeled data, the collection of unlabeled WiFi CSI data on a large scale remains challenging currently.
Therefore, we aim to examine the impact of the size of the unlabeled source dataset on the model's performance on the target dataset.

\begin{figure*}
     \centering
     \begin{subfigure}[b]{0.49\textwidth}
         \centering
         \includegraphics[width=\textwidth]{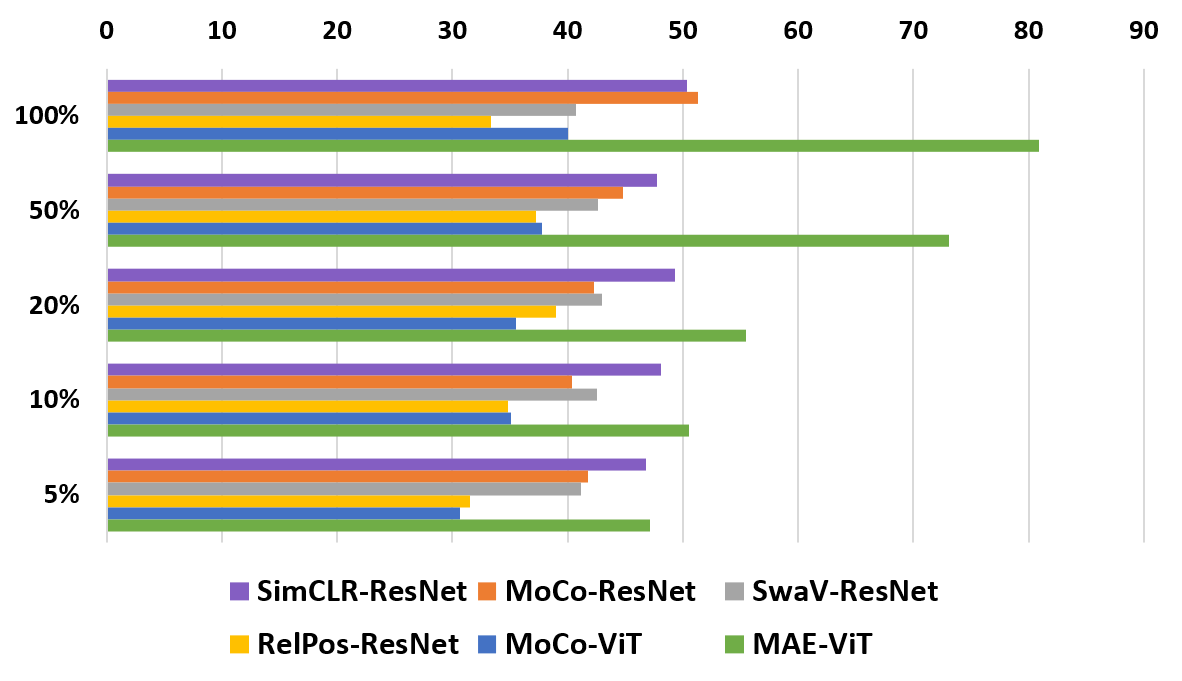}
         \caption{Linear Evaluation.}
         \label{fig:unlabeled_linear_a}
     \end{subfigure}
     \hfill
     \begin{subfigure}[b]{0.49\textwidth}
         \centering
         \includegraphics[width=\textwidth]{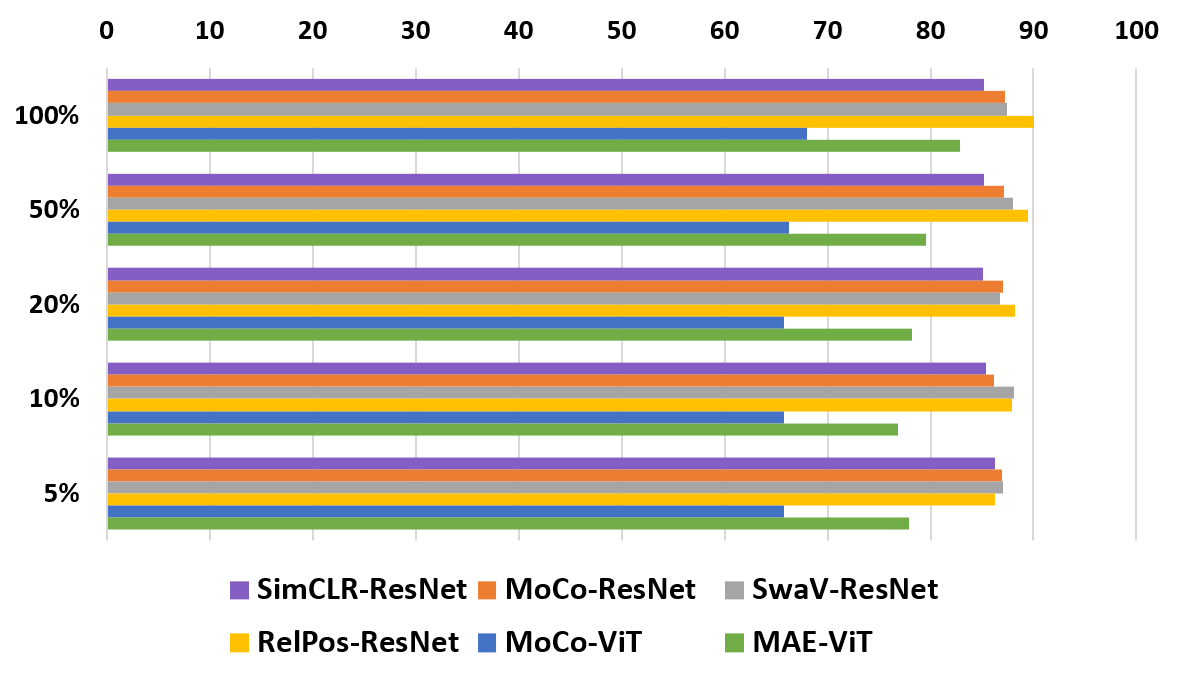}
         \caption{Fine-tuning.}
         \label{fig:unlabeled_ft_b}
     \end{subfigure}
        \caption{Analysis of Unlabeled Data Efficiency. Quantitative accuracy with varying sizes of the source training dataset. The model was trained on the source dataset collected from User 1 on Receiver 2 in Room 1, while the target dataset was collected in Room 2. (a) and (b) depict the results of linear evaluation and fine-tuning, respectively.}
        \label{fig:limitedunlabeled}
\end{figure*}

The user, receiver, and room settings of the source and target datasets remain consistent with Section \ref{sec:4.4.1.limited label data}. However, in this section, we systematically decrease the size of the source unlabeled dataset from 100\% to 5\%.
The bar charts in Figure \ref{fig:limitedunlabeled} illustrate the performance outcomes of all SSL algorithms.

From Figure \ref{fig:unlabeled_linear_a}, it is evident that the performance of MAE deteriorates as the size of the source unlabeled dataset decreases in the linear evaluation setting. However, instance discrimination, cluster discrimination, and relation prediction methods exhibit less sensitivity to the decrease in the size of the source unlabeled dataset in linear evaluation.
Figure \ref{fig:unlabeled_ft_b} demonstrates that SSL methods such as instance discrimination, cluster discrimination, and relation prediction have a significant advantage over MAE. Moreover, these methods are not sensitive to the decrease in the size of the source unlabeled dataset in the fine-tuning setting.
Importantly, MAE shows reduced sensitivity to the size of the source unlabeled dataset in the fine-tuning setting compared to linear evaluation. This can be attributed to the labeled dataset used for fine-tuning the parameters of the entire model.

\subsubsection{Section Summary}
In summary, the experiments carried out in this section provide valuable insights. 

Firstly, MAE demonstrates a high sensitivity to changes in both the quantity of labeled and unlabeled data. In contrast, algorithms such as instance discrimination, cluster discrimination, and relation prediction exhibit less sensitivity to variations in the amount of unlabeled data. 
However, their performance in the linear evaluation setting is relatively modest, indicating that the learned features from these SSL methods may not directly contribute to the classification task.
Therefore, we recommend that researchers prioritize exploring the potential of MAE and allocate more attention to harnessing its capabilities in this area.

On the contrary, in the fine-tuning setting, these algorithms demonstrate significant improvements compared to the linear evaluation. Moreover, SSL algorithms consistently outperform the supervised baseline in various fine-tuning settings, with the exception of the scenario where the labeled dataset size is only 10\%.
These findings indicate that across different labeled dataset size settings, SSL algorithms exhibit superior performance compared to the supervised baseline.
This suggests that practitioners should consider opting for self-supervised learning algorithms, particularly in cases of data scarcity.

Moreover, it is crucial to acknowledge that although SSL algorithms can outperform the supervised baseline when faced with limited labeled datasets, the magnitude of improvement is not substantial. This implies that there is untapped potential to enhance the efficiency of SSL algorithms by leveraging labeled data in our envisioned scenario. It is imperative to pursue further refinements and advancements to maximize the performance gains achieved through SSL and optimize the utilization of labeled data.

\section{Discussion and Future Work}
There are several limitations inherent in this paper. Firstly, a major challenge is the need for more available data. Compared to the extensive public datasets commonly employed in computer vision, datasets specific to CSI HAR are considerably smaller in scale. For instance, while datasets like ImageNet and CIFAR-10 contain over a million and 60,000 samples respectively, the Instagram dataset used for unlabeled pre-training boasts billions of samples \cite{mahajan2018exploring}. In contrast, the largest publicly accessible labeled dataset in the CSI field, namely Widar, comprises only 45,000 instances of action data, with each instance comprising six receiver records. The issue of insufficient data arises from two primary factors. Firstly, WIFI CSI data possesses a certain degree of privacy sensitivity, making it inaccessible for researchers and practitioners to obtain from sources such as Internet platforms akin to image-based datasets on Instagram. Secondly, labeling CSI data presents additional challenges as it requires simultaneous data collection and annotation, making the acquisition of labeled CSI datasets exceptionally arduous. Given the substantial costs associated with data collection, researchers in the CSI HAR research community should consider directing their efforts toward constructing large-scale unlabeled datasets in future endeavors.

Secondly, a limitation is the insufficient diversity of data. This issue is closely related to the scarcity of available data mentioned earlier. The largest publicly accessible CSI dataset, WIDAR, includes only 17 users and 3 rooms, significantly smaller than computer vision datasets that typically involve tens of thousands of users. Additionally, the current datasets rely on a small number of users who repeatedly perform the same actions, resulting in inherent sample repetition within the same user-action context. Consequently, the dataset lacks the necessary diversity across environmental conditions, user demographics, and action patterns. Models trained with the dataset may struggle to effectively generalize to novel environments and users within the deep learning framework. To address this limitation, it is imperative to construct unlabeled datasets exhibiting rich diversity across various factors. This will facilitate the development of more robust and generalizable models in the future.

Thirdly, this paper focuses solely on single-person activity datasets. Although previous studies have explored action recognition in multi-person scenarios, our research intentionally restricts its focus to single-person activity recognition. This deliberate choice allows us to systematically and comprehensively investigate SSL in CSI-based HAR within a well-defined research context. As a result, our study elucidates the specific research objectives related to single-person activity recognition. However, we recognize the significance of extending this research to address SSL challenges in multi-person scenarios within the CSI-based HAR domain. Therefore, future research endeavors can explore the application of SSL techniques to address the complexities and subtleties associated with recognizing activities involving multiple individuals in CSI-based HAR settings.

\section{Conclusion}
In this article, we have conducted a comprehensive investigation into the quality of features derived from four major categories of SSL methods in the context of the CSI-based HAR. 
It is important to note that although we concentrate on the CSI-based HAR task, our study has the potential to be extended to other contactless sensor applications.
We have assessed the robustness of these methods to domain transfer and their performance across varying data sizes. Our findings offer valuable insights in several key areas. Firstly, we emphasize the untapped potential of MAE as a powerful SSL algorithm, which has received relatively little attention in the research community. Secondly, we have identified opportunities for improving SSL algorithms in scenarios involving room and user transfer, where their performance can be enhanced. Further refinements are necessary to ensure their adaptability in such dynamic settings.

Furthermore, our experiments have revealed that the current SSL algorithms fail to meet the expected level of performance when the dataset is extremely limited. Their performance does not significantly exceed that of the supervised baseline, even with a small amount of labeled data. This highlights the necessity for ongoing research and development efforts to fully unleash the potential of SSL in scenarios with minimal labeled data.

While our research has provided valuable insights into SSL methods for CSI-based HAR, we acknowledge certain limitations in our study. This work serves as an initial version of research in this direction, establishing the foundation for future advancements. Our ultimate goal is to establish a publicly accessible benchmark that can be regularly updated to compare the performance of existing and future SSL algorithms across established problem domains and emerging datasets, including challenging scenarios such as multi-person scenarios. This benchmark will serve as a valuable resource for SSL researchers and practitioners in the CSI-based HAR field, raising awareness of relevant issues and aiding in selecting appropriate SSL algorithms.

In summary, this article represents a significant step in exploring SSL methods for CSI-based HAR, paving the way for further research and the development of enhanced algorithms.

\bibliographystyle{unsrt}
\bibliography{references}

\appendix
\section{Data Augmentation}
\label{App:Data Augmentation}
Our experiments involve three methods (SimCLR, MoCo, and SwAV) that require a augmentation function to implement a data augmentation module. In this section, we conduct a systematic study on the impact of using different augmentation functions for the pre-training of these methods to determine the most suitable augmentation function for each algorithm and the most general and effective augmentation function for the CSI dataset. We perform a linear evaluation on the Widar\_U123R2 dataset using Causal Net for these three SSL algorithms. We apply grid searching for hyperparameter optimization of the augmentation function and conduct our experiments under the best hyperparameter settings.

\begin{figure*}[!htbp]
\centering
\includegraphics[width=\linewidth]{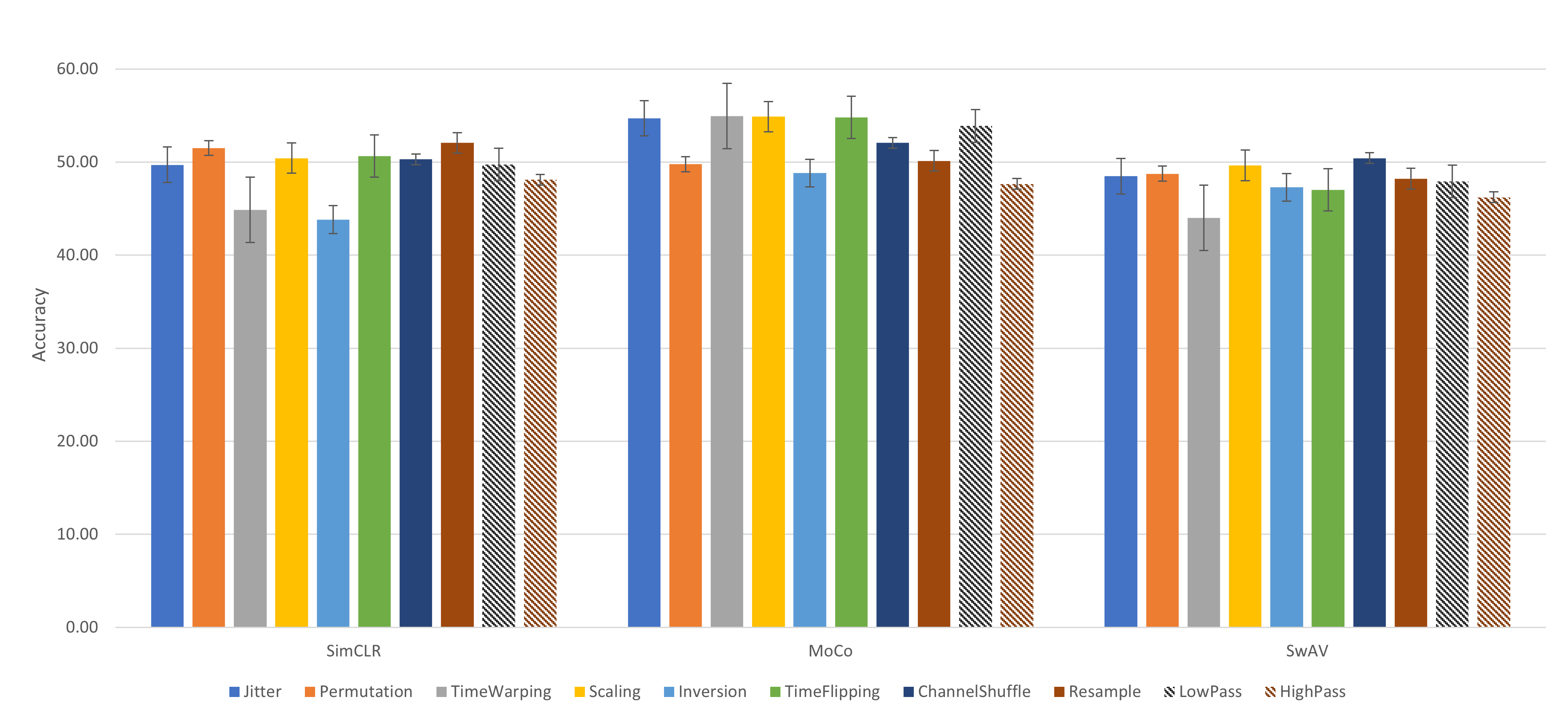}
\caption{Impact of different augmentation functions on the pre-training of SimCLR, MoCo (ResNet), and SwAV. The solid bars represent the time-domain augmentation function, while the bars filled with diagonal stripes indicate the frequency-domain augmentation function.}
\label{fig:1}
\end{figure*}

We have examined a total of ten augmentation functions in both the time and frequency domains. All of these functions have been extensively utilized in previous studies on wearable HAR \cite{tang2020exploring,qian2022makes}. Detailed information about the augmentation functions is provided in Table \ref{table:6}. Figure \ref{fig:1} illustrates the augmentation functions studied in this section.
The solid bars in Figure \ref{fig:1} represent the accuracy of augmentation functions in the time domain, while the hollow bars represent the accuracy of augmentation functions in the frequency domain. The y-axis represents the accuracy rate, while the x-axis represents the three SSL algorithms used in this section.

Firstly, regarding the SimCLR algorithm, Resample and Permutation have achieved the optimal and suboptimal results, respectively, with small variances. For the MoCo algorithm, TimeWarping and Scaling have achieved the top two results. However, TimeWarping exhibits significant variance in performance, despite its overall effectiveness. Finally, concerning the SwAV algorithm, ChannelShuffle and Scaling have obtained the best results.

Secondly, we analyze the augmentation functions in the frequency domain. It is observed that using LowPass to eliminate high-frequency components leads to better performance compared to HighPass. This finding aligns with our understanding that human activities are predominantly low-frequency signals, while high frequencies primarily represent noise.

Thirdly, we find that Scaling and ChannelShuffle consistently achieve good results across all three SSL algorithms. However, ChannelShuffle demonstrates lower performance variance, indicating more stable performance. Therefore, we recommend practitioners to choose ChannelShuffle as the augmentation function when they are uncertain about which one to employ.
\begin{table*}[!htbp]
\centering
\begin{tabular}{lll}
\toprule
Domain                     & Augmentation   & Implementation Details                                                                                                                          \\
\midrule
\multirow{8}{*}{Time}      & Jitter         & add a randomly generalized noise signal with a mean of 0 and standard deviation of 1                                                            \\
                           & Permutation    &\makecell[l]{randomly split each signal into 3 segments in the time scale, then permute\\ the segments and combine them into original shape}\\
                                               
                           & TimeWarping    & \makecell[l]{stretch and warp each signal in the temporal dimension with an arbitrary\\ cubic spline}\\
                           & Scaling        & \makecell[l]{scale each channel of the signal by a random factor that is drawn from a normal\\ distribution of mean being 1.0 and standard deviation being 0.1} \\
                           & Inversion      & multiply the value of the sample by a factor of -1\\
                           & TimeFlipping   & reverse the signal in time dimension \\
                           & ChannelShuffle & randomly permute the channels of the sample  \\
                           & Resample       & \makecell[l]{up-sample the signal in time axis to 3 times its original time steps by linear\\ interpolation and randomly down-sample to its initial dimensions }\\
\midrule
\multirow{2}{*}{Frequency} & LowPass        & use low-pass filtering to remove high-frequency components  \\
                           & HighPass       & using high-pass filtering to remove low-frequency components \\
\bottomrule
\end{tabular}
\caption{Details of Augmentation Functions in the Experiments.}\label{table:6}
\end{table*}

\section{Performance of different combinations of CSI amplitude and phase}
\label{App: different combination}
In the context of supervised learning-based CSI-HAR, researchers have explored different combinations of CSI amplitude and phase, including amplitude, phase, and amplitude-phase, to examine their effects on model performance. However, the impact of these different combinations on SSL algorithms remains uncertain. To address this gap, this section aims to investigate the influence of varying combinations of CSI amplitude and phase on SSL algorithms.
We employed the Widar\_R2 dataset to conduct our experiments, ensuring that all experiments were carried out using the optimal hyperparameters for linear evaluation.
The CSI data encompasses both amplitude and phase components, where the amplitude data is formatted as [T,30,3].
In order to process the amplitude-phase CSI data, which includes both amplitude and phase components, we modified the data format from [T,30,3,2] to [T,30,6] prior to inputting it into the model.

The results obtained in this study are presented in Tables \ref{table:7}-\ref{table:9}. In Table \ref{table:7}, we utilize the Causal Net architecture, and the findings indicate that the supervised baseline model achieves the highest performance when trained solely on the amplitude component of the CSI data. In contrast, SSL algorithms such as SimCLR and SwAV exhibit superior performance when trained on amplitude-phase data, while MoCo performs better when trained on phase data. It is noteworthy that when utilizing only the phase component of the CSI data, the performance of the supervised baseline algorithm degrades in comparison to the other two combinations. However, the performance of SSL algorithms on the phase and amplitude-phase data is comparable.

Proceeding to Table \ref{table:8}, which presents the results obtained using the ResNet encoder architecture. Similar to the findings in Table \ref{table:7}, the supervised baseline achieves the highest performance with amplitude data, while its performance is the poorest with phase data. However, all SSL algorithms exhibit a consistent pattern where the performance is strongest with amplitude-phase data and weakest with phase data.

Lastly, Table \ref{table:9} displays the results obtained when employing the ViT. The findings indicate that the supervised baseline performs the best when trained on the amplitude component of the CSI data and the worst when trained on phase data. Regarding the SSL algorithms, MoCo demonstrates the highest performance when trained on amplitude data, whereas MAE attains the best results with amplitude-phase data.

Due to the presence of significant phase noise and the absence of noise removal during training, in contrast to prior studies, it is evident that the supervised baseline model employing the phase component consistently demonstrates the poorest performance across all encoder architectures.
Remarkably, despite the subpar performance of the phase component, SSL algorithms achieve the best results when trained using amplitude-phase data. This observation implies that SSL algorithms possess the ability to extract valuable information from the noisy phase component, leading to an enhanced representation space.
These findings suggest that practitioners should adopt the amplitude-phase combination in real-world applications when utilizing SSL algorithms, as it offers a superior representation space compared to utilizing the amplitude or phase components individually. Moreover, it underscores the potential of SSL algorithms in effectively handling noisy phase data and extracting significant information from signals contaminated with noise.

\begin{table}[t]
    \begin{subtable}{.48\linewidth}
        \centering
        \caption{}{
        \begin{tabular}{l|lll}
        \toprule
        \multicolumn{1}{c|}{\multirow{2}{*}{Algorithm}} & \multicolumn{3}{c}{Combinations}                              \\
                        & \multicolumn{1}{l}{amplitude}    & \multicolumn{1}{l}{phase}                   & \multicolumn{1}{l}{amplitude-phase}       \\
        \midrule
        Supervised                           & \textbf{85.4785}        & 53.957                  & 66.541                \\
        SimCLR                               & 28.722                  & 33.149                  & \textbf{33.448}       \\
        MoCo                                 & 33.171                  & \textbf{34.034}         & 33.857                \\
        SwAV                                 & 30.293                  & 34.599                  & \textbf{34.887}       \\
        Rel-Pos                              & -                       & -                       & -                     \\
        MAE                                  & -                       & -                       & -                     \\
        \bottomrule
        \end{tabular}
        }
    \label{table:7}
    \end{subtable}

    \begin{subtable}{.48\linewidth}
        \centering
        \caption{}{
        \begin{tabular}{l|lll}
        \toprule
        \multicolumn{1}{c|}{\multirow{2}{*}{Algorithm}} & \multicolumn{3}{c}{Combinations}  \\ 
                                   & amplitude               & phase                   & amplitude-phase \\ 
        \midrule
        Supervised                                       & \textbf{94.366}         & 51.3                    & 73.027 \\ 
        SimCLR                                           & 40.432                  & 31.455                  & \textbf{41.251}          \\ 
        MoCo                                             & 41.129                  & 26.674                  & \textbf{41.173}          \\ 
        SwAV                                             & 36.392                  & 34.079                  & \textbf{37.277}          \\ 
        Rel-Pos                                & 32.175                  & 28.179                  & \textbf{32.308} \\ 
        MAE                                            & -                       & -                       & -     \\ 
        \bottomrule
        \end{tabular}
        }
    \label{table:8}
    \end{subtable} 
    \medskip
        \begin{subtable}{.48\linewidth}
        \centering
        \caption{}{
        \begin{tabular}{l|lll}
        \toprule
        \multicolumn{1}{c|}{\multirow{2}{*}{Algorithm}} & \multicolumn{3}{c}{Combinations}  \\ 
                                  & amplitude               & phase                   & amplitude-phase   \\
        \midrule
        Supervised                                       & \textbf{86.331}         & 42.247                  & 82.35635  \\
        SimCLR                                           & -                       & -                       & -         \\
        MoCo                                             & \textbf{62.811}         & 50.327                  & 57.609    \\
        SwAV                                             & -                       & -                       & -         \\
        Rel-Pos                                & -                       & -                       & -                   \\
        MAE                                              & 69.242                  & 51.356                  & \textbf{79.148}          \\
        \bottomrule
        \end{tabular}
        }
    \label{table:9}
    \end{subtable} 
\caption{The performance of the algorithms on different combinations of CSI amplitude and phase using 
(a) Causal Net 
(b) ResNet
(c) ViT
is presented. The best accuracy achieved by a self-supervised learning algorithm is denoted in bold. A "-" indicates that the algorithm cannot utilize Causal Net as the encoder architecture.
} 
\end{table}

\end{document}